\documentclass[twocolumn,tighten]{aastex7}
\usepackage[utf8]{inputenc}
\usepackage[T1]{fontenc}
\usepackage{array}
\usepackage{amsmath}
\usepackage{mathtools}
\usepackage{graphics,graphicx}
\usepackage{enumitem}
\usepackage{natbib}
\usepackage[tight-spacing=true,table-number-alignment=center,table-align-text-post=false,table-align-exponent=false]{siunitx}
\usepackage{color}
\usepackage[nameinlink,capitalize]{cleveref}
\usepackage[format=hang,font=small,labelfont=bf,figurename=Fig.]{caption}

\begin{document}

\title[TIRAMISU: Non-LTE radiative transfer for molecules in exoplanet atmospheres]{TIRAMISU: Non-LTE radiative transfer for molecules in exoplanet atmospheres}

\author[orcid=0000-0002-5941-9936]{Charles A. Bowesman}
\affiliation{Department of Physics and Astronomy, University College London, Gower Street, WC1E 6BT London, UK}
\email[show]{charles.bowesman@ucl.ac.uk}
\author[orcid=0000-0001-9286-9501]{Sergei N. Yurchenko}
\affiliation{Department of Physics and Astronomy, University College London, Gower Street, WC1E 6BT London, UK}
\email{s.yurchenko@ucl.ac.uk}
\author[orcid=0000-0003-2241-5330]{Ahmed Al-Refaie}
\affiliation{Department of Physics and Astronomy, University College London, Gower Street, WC1E 6BT London, UK}
\email{ahmed.al-refaie@ucl.ac.uk}
\author[orcid=0000-0002-4994-5238]{Jonathan Tennyson}
\affiliation{Department of Physics and Astronomy, University College London, Gower Street, WC1E 6BT London, UK}
\email{j.tennyson@ucl.ac.uk}



\begin{abstract}

The TIRAMISU code, a new program for computing on-the-fly non-LTE molecular spectra and opacities for solving self-consistent radiative transfer problems in exoplanet atmospheres, is presented.
The ultra-hot Jupiter KELT-20\,b is used as a case study to identify the wavelength regions at which non-LTE effects may be detectable.
It is shown that upper atmospheric OH in vibrational non-LTE should be observable primarily via hot bands in the mid-infrared and enhanced photodissociation in the visible.
Varying the abundance of OH in non-LTE demonstrates a non-linear relationship between the abundance and the strength of non-LTE effects.
Using recent calculations of the photodissociation probabilities of OH it is shown that non-LTE effects can increase the total photodissociation rate by two orders of magnitude in the upper atmosphere, which is likely to have a significant impact on atmospheric and chemical modelling.
Increases and reductions in the molecular opacities under non-LTE conditions may lead to the mischaracterisation of molecular abundances in retrievals that only consider opacities computed under LTE.
Collisional data requirements to support future non-LTE modelling for a variety of exoplanet atmospheres and across a wide range of wavelengths are discussed.

\end{abstract}


\section{Introduction}

While the ability to characterise exoplanet atmospheres continues to be enhanced by the development of new instruments and analysis techniques, many studies still rely on molecular opacity data computed under the assumption of local thermodynamic equilibrium (LTE) conditions.
Non-LTE effects, specifically where the quantum mechanical energy level populations of a molecule deviate from those given by the Boltzmann distribution, are known to be important for modelling observed spectral features in Earth's atmosphere \citep{12FeKu.nLTE} which are in turn important for modelling its structure \citep{01LoTa.nLTE}.
Elsewhere in the solar system, non-LTE effects are known to be important in the atmospheres of Venus \citep{07LoDrCa.nLTE,09GiLoDr.nLTE}, Mars \citep{76JoBeMc.nLTE,16PiLoMa.nLTE}, Saturn \citep{11TaBaFu.planets}, Jupiter \citep{24SaLoGa.CH4}, Uranus \citep{04EnLeDr.planets} and Neptune \citep{10FlDrBu.planets}, as well as in other objects such as Titan \citep{11GaLoFu.CH4,13ReKuFa.nLTE} and Pluto \citep{25LeWoLa.nLTE}.
Beyond this, non-LTE effects are well known to be important in environments from cool molecular clouds \citep{09LivaKl.nLTE} to stellar photospheres \citep{05Asplund.nLTE}.
Given the ubiquity of non-LTE effects in the solar system, it would stand to reason that they would likely play a role in the atmospheres of exoplanets. 
Indeed, such effects are increasingly being considered in the modelling of exoplanet atmospheres \citep{21FoYoSh.nLTE,21BoFoKo.nLTE}, though many studies have focussed on their application to atomic spectra \citep{19FiHexx.nLTE,20YoFoKo.exo,25FoSrKo.nLTE} or have been limited to rotational non-LTE \citep{23ChBiCa.nLTE}.
The non-LTE problem is more complex for molecules owing to the considerably larger number of energy levels.
The number of radiative transitions between these levels is therefore much greater; comprehensive knowledge of transition probabilities is essential for modelling the radiative processes that inform the non-LTE statistical equilibrium.
Methods of simplifying non-LTE calculations exist, such as the Treanor method that instead parametrises the Boltzmann distribution using two decoupled rotational and vibrational temperatures \citep{68TrRiRe.nLTE}.
Non-LTE effects arising from such a distribution have been shown to significantly impact the modelled absorption spectra of several molecules believed to be important in exoplanet atmospheres \citep{22WrWaYu.nLTE}.
In practice however, neither the rotational nor vibrational distributions need to be Boltzmann-like in non-LTE.

The ExoMol project \citep{jt939} provides a database of atomic and molecular line list data, as well as cross sections and k-tables pre-computed under LTE conditions \citep{jt801}.
As part of this project, a new line list for the hydroxyl radical (OH) was recently calculated, refined against laboratory measurements such that it reproduces experimental transition frequencies in the observed bands \citep{jt969}.
A study of the photodissociation dynamics of OH in its ground X~$^{2}\Pi$ state was also performed, considering three different dissociation channels \citep{jt982}.
This allows for the computation of temperature and pressure dependent photodissociation cross sections of OH.
As with absorption cross sections arising from bound-bound transitions, the magnitude of photodissociation or continuum absorption cross sections are dependent on the initial state populations.
This means they will vary as a function of temperature under both LTE and non-LTE conditions; enhanced excited state populations in non-LTE environments can considerably increase the photodissociation cross section and in turn the photodissociation rate.
These rates are important for modelling the chemical equilibria of exoplanet atmospheres and can have a significant impact on the derived structure and composition, even if the molecule is not directly observed.

The solution to the non-LTE statistical equilibrium equation is implicitly coupled to the radiative transfer problem, owing to the dependence of the rates of stimulated emission and absorption on the radiation in the media.
Many methods have been developed to solve this non-linear problem, most of which rely on linearisation techniques \citep{69AuMi3.nLTE}.
Perhaps the most prominent is the method of multilevel approximate Lambda iteration (MALI) of \citet{85ScCa.nLTE} and \citet{91RyHuxx.nLTE,92RyHuxx.nLTE}.
This method employs the Lambda operator splitting of \citet{73Cannon.nLTE} to apply preconditioning terms to the statistical equilibrium equations, significantly accelerating the convergence of the method.
This preconditioning also circumvented the main failure of regular Lambda iteration, which was the pseudo-convergence to false minima due to the trapping of photons in optically thick media \citep{15LaPaJo.nLTE}.
\citet{95TrFaxx.nLTE} showed that application of a Gauss-Seidel (GS) iteration scheme, rather than the Jacobi-like iteration of MALI, yielded improved convergence speed and was more resilient in environments highly deviated from LTE.
\citet{06AsTr.nLTE} provide a detailed overview of these methods.

Here the non-LTE radiative transfer code TIRAMISU (Tridiagonal Integrated Radiative transfer And Molecular opacity Iterative Solver Utility) is presented with a case study of its application to non-LTE OH in the atmosphere of an ultra-hot Jupiter.
The code is designed for solving the coupled radiative transfer and statistical equilibrium problem in exoplanet atmospheres, comprising an arbitrary mixture of atomic and molecular opacity sources and with full consideration of the cross-coupling between overlapping opacity sources in the same and adjacent regions of the atmosphere.
Section 2 outlines the method for solving the non-LTE radiative transfer problem.
Section 3 details the data available for modelling the rates of collisional processes for OH.
Section 4 describes the model atmosphere constructed for testing the new code and section 5 gives the results obtained from modelling OH in this example atmosphere.

\section{Method}
\label{sec:method}

It has long been understood that accurate modelling on non-LTE level populations requires the solution of the coupled radiative transfer problem, owing to the dependence of the rates of stimulated processes on the background radiation.
As this radiation changes as a function of the level populations, the problem is solved iteratively until a solution is converged upon.
To obtain a given set of non-LTE populations, one must solve a statistical equilibrium equation describing the balance between the processes that populate and depopulate a given energy level.
These are then used to compute absorption and emission cross sections for the various atmospheric species.
A formal solution to the radiative transfer problem is then obtained and the results fed back into the next iteration.
A range of methods have been developed over the years to optimise the speed and accuracy of this convergence.
Here, a GS iteration scheme as developed by \citet{95TrFaxx.nLTE} and \citet{97FaTrAu.nLTE} is employed in conjunction with successive over-relaxation to reduce the total number of iterations.
This section details the technical choices that were made in the development of the TIRAMISU code, as there are a variety of implementations available in the literature.
Portions of the code used to construct the model atmosphere and process the LTE opacity sources was adapted from the TauREx 3 source code \citep{TauREx3,22AlChVe.exo}.

\subsection{Statistical equilibrium}

The statistical equilibrium equation for level populations can be constructed in numerous ways depending on the processes under consideration.
Here, a comprehensive formulation comprising radiative, collisional, and chemical rates is utilised such that
\begin{equation}
    \label{eq:statistical_equilibrium}
    \begin{aligned}
    \frac{dn_{i}}{dt} = &\sum_{j}\left(n_{j}C_{ji} - n_{i}C_{ij}\right) + F_{i} - n_{i}D_{i}\\
    & + \int_{0}^{\infty}d\nu\oint_{4\pi}d\Omega\,\frac{4\pi}{h\nu}\bigg(n_{j}U_{ji} - n_{i}U_{ij} \\
    & + \left(n_{j}V_{ji} - n_{i}V_{ij} - n_{i}V_{ic} + n_{c}V_{ci}\right)I^{*}_{\nu} \\
    & + \chi^{\dagger}_{ji}\sum_{kl}\Lambda^{*}\left[n_{k}U_{kl}/\chi^{\dagger}_{\nu}\right]\bigg) = 0,
    \end{aligned}
\end{equation}
where the $C_{ij}$ terms describe the collisional rates between levels and $F_{i}$ and $D_{i}$ describe the chemical formation and destruction rates, respectively.
The radiative rates are described by $U_{ij}$ and $V_{ij}$, where $U_{ij}$ describes the rate of spontaneous emission
\begin{equation}
    \label{eq:u_ij}
    U_{ij} = \begin{cases}
        \frac{h\nu}{4\pi}A_{ij}\phi_{ij}\left(\nu\right)& \hfill\textrm{if } E_{i} > E_{j}\\
        0 & \hfill\textrm{otherwise,}
    \end{cases}
\end{equation}
and $V_{ij}$ the rate of stimulated processes, such that
\begin{equation}
    \label{eq:v_ij}
    V_{ij} = \frac{h\nu}{4\pi}B_{ij}\phi_{ij}\left(\nu\right).
\end{equation}
The quantity $\phi_{ij}$ describes the integral normalised line profile.
This form of the statistical equilibrium equations was outlined by \citet{92RyHuxx.nLTE} and describes the fully preconditioned rates, building on the 
method that only preconditioned ``within'' the profile of each transition, as given by their earlier work \citep{91RyHuxx.nLTE}.
The term $I^{*}_{\nu}$ is the preconditioned intensity and is given by
\begin{equation}
    \label{eq:preconditioned_intensity}
    I^{*}_{\nu} = \Lambda\left[\eta^{\dagger}_{\nu}/\chi^{\dagger}_{\nu}\right] - \Lambda^{*}\left[\sum_{ij}\eta^{\dagger}_{ij}/\chi^{\dagger}_{\nu}\right],
\end{equation}
where $\Lambda$ and $\Lambda^{*}$ are the full and approximate Lambda operators, $\eta_{\nu}$ is the emissivity and $\chi_{\nu}$ is the absorption.
A primer on these operators can be found in \citet{14HuMi.nLTE} though detailed discussion of their application to this work is given in Section \ref{subsec:construiction_of_lambda_operator}.
The superscript ${\dagger}$ is used to denote a quantity evaluated using results from the previous iteration, when available.
It is worth noting that the approximate Lambda operator in \cref{eq:preconditioned_intensity} acts only on the total sum of all emissivity from the molecule in question, though still considers the total opacity from all atmospheric constituents.
The emissivity is defined here as
\begin{equation}
    \label{eq:eta_ij}
    \eta_{ij} = \frac{h\nu}{4\pi}n_{i}A_{ij}\phi_{ij}\left(\nu\right),
\end{equation}
and the absorption as
\begin{equation}
    \label{eq:chi_ij}
    \chi_{ij} = \frac{h\nu}{4\pi}\left(n_{j}B_{ji}\phi_{ji}\left(\nu\right) - n_{i}B_{ij}\phi_{ij}\left(\nu\right)\right).
\end{equation}

Preconditioning in this manner aims to ensure that the set of populations obtained have physical values, such as ensuring that they are positive \citep{97SoTr.nLTE}.
Linearisation in this manner also offers a performance increase, particularly in optically thick media \citep{94AuFaTr.nLTE}.
In \cref{eq:u_ij,eq:v_ij}, $A_{ij}$ is the Einstein A coefficient describing the rate of spontaneous emission from the $i$-th to $j$-th level and $B_{ij}$ is the Einstein B coefficient describing the rate of stimulated processes; when $E_{i} > E_{j}$ this is stimulated emission and when $E_{i} < E_{j}$ this is absorption.
$B_{ic}$ describes the continuum absorption or photodissociation rate, where the subscript ``c'' denotes a continuum level; $B_{ci}$ describes the photo-attachment rate which depends on $n_{c}$.
Here, $n_{c}$ is used to denote the population of the limiting species in the photo-attachment process, i.e.: for OH in a Hydrogen-dominated atmosphere, photo-attachment is limited by the number of free Oxygen atoms. 

The Einstein B coefficient, $B_{ij}$, can be related to the Einstein A coefficient but the exact expression used varies in the literature depending on whether $B_{ij}$ is being used to describe a rate dependent on either the intensity, irradiance or energy density incident on the atom or molecule \citep{82Hilbor.nLTE}.
Here, the Einstein B coefficient describing the process of stimulated emission is 
\begin{equation}
    \label{eq:einstein_b_ul}
    B_{ij} = \frac{A_{ij}c^{2}}{2h\nu_{ij}^{3}},
\end{equation}
where $h$ is the Planck constant, $c$ is the speed of light in vacuum and $\nu_{ij}$ is the transition frequency \citep{78Mihala.nLTE}.
This can be related to the probability of absorption through the expression
\begin{equation}
    \label{eq:einstein_b_lu}
    B_{ji} = B_{ij} \frac{g_{i}}{g_{j}},
\end{equation}
where $g_{i}$ and $g_{j}$ are the degeneracies of the $i$-th and $j$-th levels.
This form of the Einstein B coefficient is expressed in terms of a rate per radiation density in units of s$^{-1}$m$^{2}$J$^{-1}$.
This term alone does not provide a direct rate for the stimulated processes without also considering the radiation field in the medium to drive such processes, i.e.: the energy density in J~m$^{-2}$.
This incident energy density is accounted for in the preconditioned intensity terms defined by \cref{eq:preconditioned_intensity}.

To evaluate the intensity driving stimulated processes, the atmosphere is divided into discrete layers in order to solve the radiative transfer problem.
The intensity $I_{\Omega,\nu,l}$ at each layer of the atmosphere can be computed as
\begin{equation}
    \label{eq:intensity_lambda}
    I_{\Omega,\nu,l} = \mathbf{\Lambda}_{\Omega,\nu,l} \left[S_{\Omega,\nu,l}\right],
\end{equation}
where $\mathbf{\Lambda}_{\Omega,\nu,l}$ is the so-called Lambda operator, $S_{\Omega,\nu,l}$ is the Source function and all quantities are dependent on the direction of radiation $\Omega$, frequency $\nu$ and an index $l$ which tracks the layer of the atmosphere such that $l = 0$ is the bottom of the atmosphere.
The Source function is expressed as
\begin{equation}
    \label{eq:source_function}
    S_{\Omega,\nu,l} \equiv \frac{\eta_{\Omega,\nu,l}}{\chi_{\Omega,\nu,l}},
\end{equation}
where $\chi_{\Omega,\nu,l}$ and $\eta_{\Omega,\nu,l}$ are the sum of the terms given by \cref{eq:chi_ij,eq:eta_ij} for all transitions of all species in each layer.
Under LTE conditions, this expression is given by the Planck function; here these terms are computed directly from the input line lists using the populations obtained at each iteration.
The Lambda operator transforms a description of the radiation travelling through a region of the atmosphere, the source function, into the  intensity at that point.
It can then be recognised from \cref{eq:intensity_lambda} that an obvious choice of the Lambda operator would be a function of the absorption or optical thickness at each layer.
Through application of the operator splitting technique, an approximate form of the operator, $\Lambda^{*}$, can be separated out which acts only on the local part of the spatial grid, i.e. within the same layer \citep{73Cannon.nLTE}.
This term describes the self-scattering component of the radiation in a given layer; photons that scatter within the same discretised region of the atmosphere do not contribute to the overall transport of radiation.
Consequently, this component is subtracted from the action of the overall radiation field in \cref{eq:preconditioned_intensity}.

It was pointed out by \citet{92RyHuxx.nLTE} that the term $\chi_{ij}$ is not linear with respect to the populations, owing to the dependence on both $n_{i}$ and $n_{j}$.
This prompted the introduction of the so-called Psi-operator which instead acts only on the emission such that
\begin{equation}
    \label{eq:intensity_psi}
    I_{\Omega,\nu,l} = \mathbf{\Psi}_{\Omega,\nu,l} \left[\eta_{\Omega,\nu,l}\right].
\end{equation}
It can be seen that the two operators are related via
\begin{equation}
    \label{eq:psi_lambda}
    \mathbf{\Psi}_{\Omega,\nu,l}\left[\cdots\right] = \mathbf{\Lambda}_{\Omega,\nu,l} \left[\left(\chi^{\dagger}_{\Omega,\nu,l}\right)^{-1}\cdots\right],
\end{equation}
where the $\chi^{\dagger}_{\Omega,\nu,l}$ term is evaluated from the previous iteration, removing the non-linear dependence on populations.
Hence the terms in \cref{eq:statistical_equilibrium,eq:preconditioned_intensity} are interchangeably quoted in terms of $\Psi^{*}\left[\eta\right]$ in the literature; the Lambda operator notation is used here as it is more widespread.
The right-hand term in \cref{eq:preconditioned_intensity} provides a weighting based on the molecule's contribution to the total intensity, which is useful when considering a medium with multiple background opacity sources.
The final summation term in \cref{eq:statistical_equilibrium} is the cross-coupling term that preconditions against the effects of overlapping transitions.

All of the quantities in \cref{eq:statistical_equilibrium,eq:u_ij,eq:v_ij,eq:preconditioned_intensity,eq:eta_ij,eq:chi_ij} have a dependence on the layer index $l$, because various properties such as the molecular number density, absorption and emission profiles and background opacities change depending on the position in the atmosphere.
The subscript was omitted in these equations for visual clarity.

\begin{figure}
    \centering
    \includegraphics[width=1.0\linewidth]{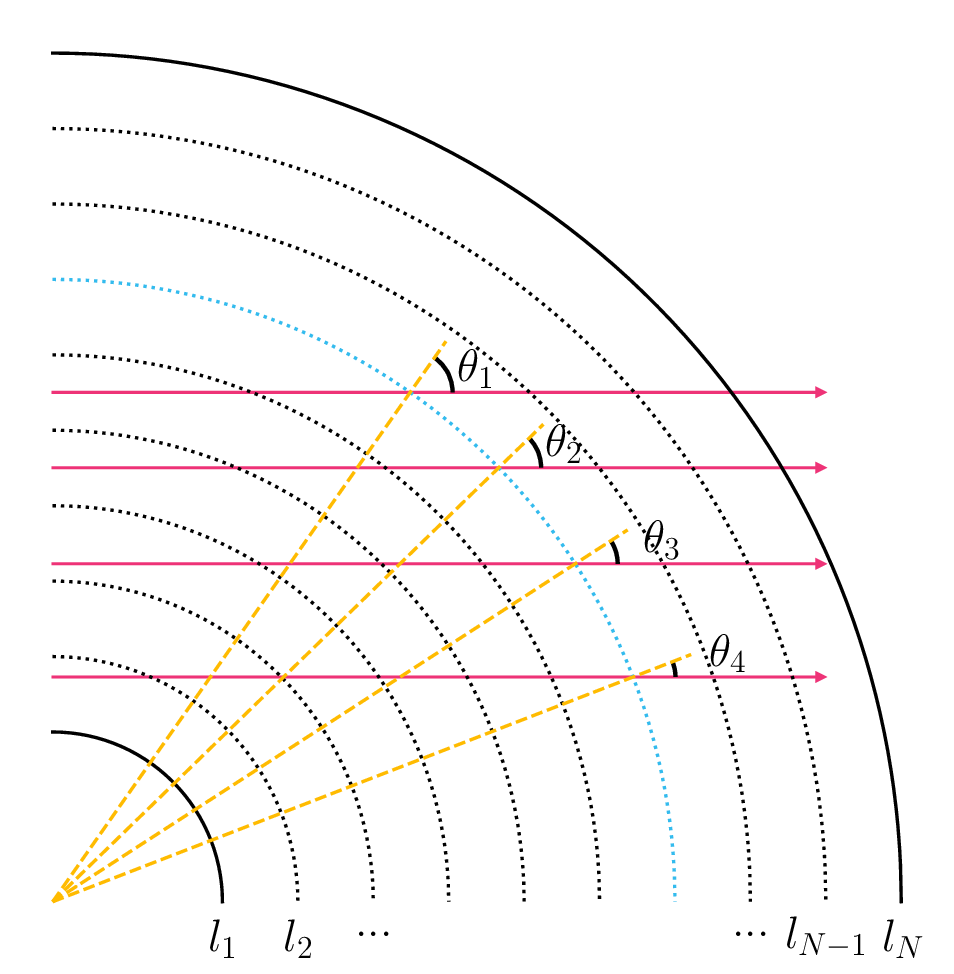}
    \caption{Geometry of a model atmosphere subdivided into $N$ layers. A set of rays (solid pink) are considered that intersect each layer and have an angle, $\theta$,  between the ray paths and the corresponding radial lines (dashed yellow). The points of intersection are chosen such that $\mu = \cos\left(\theta\right)$ correspond to the Gauss-Legendre quadrature points. Only four rays are shown here for visual clarity; 50 angular quadrature points are used in this work. Here $l_{1}$ is the surface of the planet and $l_{N}$ is the top of the atmosphere.}
    \label{fig:geometry}
\end{figure}

The angular integration in \cref{eq:statistical_equilibrium} accounts for the emission of photons in and absorptions of photons from all directions.
Considering the paths, or ``rays'', that these photons travel through the atmosphere, the integration over possible paths through the atmosphere is performed by considering the angle $\theta$ that the ray makes with the radial line.
This allows for the integration to be transformed from $d\Omega$ to $d\mu$, where $\mu = \cos\left(\theta\right)$.
The corresponding geometry is shown in \cref{fig:geometry}.
In practice, both integrals in \cref{eq:statistical_equilibrium} are performed numerically, with the integral $d\mu$ performed using Gauss-Legendre quadrature over the interval $\left[0, +1\right]$.
The integral $d\nu$ is performed using Simpson integration over the range of the line profile; binned line profiles are utilised to conserve the total intensity when computing all $\nu$-dependent quantities.

It should be noted that, while the normalised absorption and emission profiles of a single transition are equivalent under complete radiative redistribution and isotropic scattering \citep{98KuGuOg.nLTE,14HuMi.nLTE}, the normalised band profile of an aggregated vibronic transition is not.
This is primarily due to how the emission coefficients are linearly dependent on transition energy, while absorption coefficients are inversely proportional to the square of the transition energy \citep{jt708}.
While many works are able to make simplifications to \cref{eq:statistical_equilibrium} assuming the equivalencies of the line profiles $\phi_{ij}\left(\nu\right)$ in \cref{eq:u_ij,eq:v_ij} under permutation of the indices $i, j$, care must be taken here as this no such factorisation can be made.
The source function $S_{ji}$ is unchanged by permutation of $i$ and $j$, however, as these indices only correspond to a value of the source function at the $\nu_{ij}$ transition energy.
The full preconditioning strategy employed here allows for line profile overlaps, both between transitions of the same molecule and overlaps with background opacity sources.
This accounts for stimulated processes occurring in a molecule due to photons emitted by other molecules, and the equivalent for absorption.
Some models that are designed for studying environments such as the ISM do not consider line overlaps \citep{07vaBlSc.nLTE} and others introduce approximations, often only focusing on overlap with transitions arising from the same species \citep{15HsLaBe.nLTE,20Nesterenok.nLTE,24YoNoFu.nLTE}.
These approaches are appropriate for the cool environments they consider, but consideration of overlapping non-continuum opacities from different absorbers is necessary for hot exoplanet atmospheres.
Indeed, it can be seen that broad continuum features arising from photodissociation can overlap entirely with bound-bound bands.
It is also often the case that vibrational transitions within a given electronic state have considerable overlap with other transitions that have the same $\Delta v$.
Accordingly for the treatment of vibronic non-LTE, a formalism that accounts for this overlap was necessary.

Given a suitable form of $\mathbf{\Lambda_{\mu,\nu,l}}$ it is possible to compute an initial solution to \cref{eq:statistical_equilibrium}, though it can be recognised that the set of populations obtained this way can then be used to calculate new emission and absorption spectra.
The change in these quantities then also changes Source function, intensities and preconditioned terms in \cref{eq:preconditioned_intensity}, which in turn changes the populations obtained from \cref{eq:statistical_equilibrium}.
Therefore, a solution to this problem must be obtained iteratively; here a GS iteration scheme is used to find the convergent set of populations.
In this approach, radiation is propagated into the atmosphere from the top and then back out again from the bottom and a form of the Lambda operator is chosen that couples adjacent layers of the atmosphere.
During each step through the atmosphere, the value of the Source function and Lambda operator are updated for that layer.
As the value of the Lambda operator at a given layer depends on properties of its adjacent layers, layers must be evaluated sequentially \citep{18YaReLa.nLTE}.
Other iterative schemes such as MALI can be implemented such that they are able to simultaneously evaluate intensities and obtain new populations for every layer, though these methods often require a greater number of iterations to achieve convergence.

\subsection{Construction of Lambda operator}
\label{subsec:construiction_of_lambda_operator}

While many different forms of the Lambda operator have been utilised in the literature, some implementations achieve faster and more accurate convergence \citep{19Janett.nLTE}.
Adopting a short characteristics solution to the radiative transfer problem \citep{87OlKu.nLTE}, a matrix form of the approximate lambda operator $\Lambda_{\mu, \nu, l}^{*}$ that consists of the tridiagonal elements of the full Lambda operator is used here.
Accordingly, $\Lambda_{\mu, \nu, l}^{*}$ couples a given layer on the diagonal to layers physically above and below it in the atmosphere via the above- and below-diagonal elements \citep{88KuAu.nLTE}.

The elements of $\Lambda_{\mu, \nu, l}$ are determined from the interpolation coefficients used to obtain the intensity at each layer under the short characteristics scheme.
\citet{87OlKu.nLTE} detailed the construction of a tridiagonal approximate Lambda operator with a short characteristics scheme using parabolic interpolation.
More recently, an alternative B\'{e}zier interpolation scheme has been detailed \citep{13dePi.nLTE,13StBu.nLTE,19KrKoFa.nLTE,21KoKuPi.nLTE} which yields more accurate results \citep{19Janett.nLTE}; these schemes have been adapted here for use in both inward and outward propagation of radiation.
While the parabolic interpolation scheme is more widely used it only guarantees positive intensities and thereafter positive populations when only the diagonal approximate lambda operator is used \citep{91RyHuxx.nLTE} and can then still provide unphysical results if the grid is chosen poorly, resulting in large gradients \citep{88KuAu.nLTE}.
The B\'{e}zier interpolation scheme however ensures monotonicity thanks to the use of control points.
Indeed, initial testing on non-LTE OH opacities computed with parabolic interpolants often resulted in regions of negative intensity and associated negative populations, while the B\'{e}zier scheme was strictly positive.
Hence, a useful identity is defined such that
\begin{equation}
    \label{eq:delta}
    \delta^{\pm}_{l} \equiv \lvert\frac{\tau_{l}-\tau_{l\mp1}}{\mu}\rvert,
\end{equation}
where $\delta^{+}_{l}$ is the optical depth step at all frequency and angle points between the layer $l$ and the above layer $l - 1$ when propagating radiation inwards and $\delta^{-}_{l}$ is the optical depth step between the current layer and the layer below when propagating radiation outwards.
It should be noted that $\tau_{l+1} \geq \tau_{l}$ as $\tau_{l}$ is summed cumulatively in from the upper boundary of the atmosphere, implying all $\delta^{\pm}_{l} \geq 0$.
Here and in the rest of this paper, terms with superscript ``$+$'' relate to the inward propagation of radiation and those with superscript ``$-$'' relate to the outward flow.
Using this, the inward and outward directed intensities are obtained for each layer as
\begin{equation}
    \label{eq:intensity_bezier}
    \begin{aligned}
        I^{+}_{l} = &\; I^{+}_{l-1}\exp{\left(-\Delta\tau^{+}_{l}\right)} + \alpha^{+}_{l}S_{l} + \beta^{+}_{l}S^{\prime}_{l-1} + \gamma^{+}_{l}C^{+}_{l},\\
        I^{-}_{l} = &\; I^{-}_{l+1}\exp{\left(-\Delta\tau^{-}_{l}\right)} + \alpha^{-}_{l}S_{l} + \beta^{-}_{l}S^{\prime}_{l+1} + \gamma^{-}_{l}C^{-}_{l},
    \end{aligned}
\end{equation}
where $\alpha$, $\beta$ and $\gamma$ are the B\'{e}zier interpolants and are given by
\begin{equation}
    \label{eq:bezier_coefs}
    \begin{aligned}
    \alpha_{l}^{\pm} & = \frac{2 + \left(\delta^{\pm}_{l}\right)^{2} - 2\delta^{\pm}_{l} - 2\exp\left(-\delta^{\pm}_{l}\right)}{\left(\delta^{\pm}_{l}\right)^{2}},\\
    \beta_{l}^{\pm} & = \frac{2 - \left(2 + 2\delta^{\pm}_{l} + \left(\delta^{\pm}_{l}\right)^{2}\right)\exp\left(-\delta^{\pm}_{l}\right)}{\left(\delta^{\pm}_{l}\right)^{2}},\\
    \gamma_{l}^{\pm} & = \frac{2\delta^{\pm}_{l} - 4 + \left(2\delta^{\pm}_{l} + 4\right)\exp\left(-\delta^{\pm}_{l}\right)}{\left(\delta^{\pm}_{l}\right)^{2}}.\\
    \end{aligned}
\end{equation}
In cases where $\delta^{\pm}_{l} < 0.14$, the coefficients in \cref{eq:bezier_coefs} are evaluated using Horner's method, as in the PORTA code of \citet{13StBu.nLTE}.
It can be noted for all non-negative values of $\delta^{\pm}_{l}$ the coefficients of \cref{eq:bezier_coefs} are strictly positive.
$C^{\pm}_{l}$ represents a control point used to ensure monotonicity in the interpolated curve and is calculated as
\begin{equation}
    \label{eq:bezier_control_point}
    C^{\pm}_{l} = \frac{1}{2} \left(C^{\pm0}_{l} + C^{\pm1}_{l\mp1}\right),
\end{equation}
such that it is the mean of the control points associated with adjacent layers.
These further control points are calculated as
\begin{equation}
    \label{eq:bezier_c0_c1}
    \begin{aligned}
        C^{\pm0}_{l} = &\; S_{l} + \frac{\delta^{\pm}_{l}}{2}\frac{dS_{l}}{d\tau},\\
        C^{\pm1}_{l\mp1} = &\; S_{l\mp1} - \frac{\delta^{\pm}_{l}}{2}\frac{dS_{l\mp1}}{d\tau}.\\
    \end{aligned}
\end{equation}
The derivative term is computed using the expression
\begin{equation}
    \label{eq:bezier_ds_dtau}
    \frac{dS_{l}}{d\tau} = \frac{\Delta{}S_{l-1/2}\Delta{}S_{l+1/2}}{\zeta_{l}\Delta{}S_{l+1/2} + \left(1-\zeta_{l}\right)\Delta{}S_{l-1/2}},
\end{equation}
with the caveat that the derivative is taken to be zero when the numerator is negative.
$\Delta{}S_{k\pm1/2}$ represents the source function gradient as a function of optical depth between the current and adjacent layer in either direction, obtained as
\begin{equation}
    \label{eq:bezier_delta_s}
    \begin{aligned}
        \Delta{}S_{l-1/2} = &\: \frac{S_{l} - S_{l-1}}{\tau_{l}-\tau_{l-1}},\\
        \Delta{}S_{l+1/2} = &\: \frac{S_{l+1} - S_{l}}{\tau_{l+1}-\tau_{l}},\\
    \end{aligned}
\end{equation}
and
\begin{equation}
    \label{eq:bezier_zeta}
    \zeta_{l} = \frac{1}{3}\left(1 + \frac{\delta_{l}}{\delta_{l} - \delta_{l-1}}\right).
\end{equation}
While the control point $C^{\pm0}_{l}$ has a dependence on the source function at layers $l-1, l, l+1$, $C^{\pm1}_{l\mp1}$ depends on layers $l, l\pm1, l\pm2$.
Accordingly, this second control point cannot be calculated in the first two layers of the inward or outward passes and instead $C^{\pm}_{l} = C^{\pm0}_{l}$ is used.
The total intensity at each layer is simply given by the mean of the inward and outward directed intensities
\begin{equation}
    \label{eq:intensity_layer}
    I_{l} = \frac{1}{2}\left(I^{+}_{l} + I^{-}_{l}\right).
\end{equation}
As \citet{87OlKu.nLTE} did for the Parabolic interpolation coefficients, the Lambda operator for the layer $l$ can be constructed by considering the terms in \cref{eq:intensity_bezier} that act on $S_{l}$.
As a tridiagonal approximate Lambda operator is used here, only the terms that act on $S_{l}$ in the expressions for the inward and outward intensities in layers $l-1$ to $l+1$ are considered.
This yields
\begin{equation}
    \label{eq:lambda_components}
    \begin{aligned}
        \Lambda^{+}_{l} = &\; \left(\alpha^{+}_{l} + \gamma^{+}_{l}\right)\left(1 + \exp\left(-\delta^{+}_{l+1}\right)\right) + \beta^{+}_{l+1},\\
        \Lambda^{-}_{l} = &\; \left(\alpha^{-}_{l} + \gamma^{-}_{l}\right)\left(1 + \exp\left(-\delta^{-}_{l-1}\right)\right) + \beta^{-}_{l-1}.\\
    \end{aligned}
\end{equation}
The lambda operator is then constructed as the mean of the inward and outward components given in \cref{eq:lambda_components}
\begin{equation}
    \label{eq:lambda_layer}
    \Lambda_{\mu,\nu,l}^{*} = \frac{1}{2}\left(\Lambda^{+}_{l} + \Lambda^{-}_{l}\right),
\end{equation}
which can then be substituted into \cref{eq:preconditioned_intensity} to compute precondition the intensity.

All of the terms calculated in \cref{eq:delta,eq:intensity_bezier,eq:bezier_coefs,eq:bezier_control_point,eq:bezier_c0_c1,eq:bezier_ds_dtau,eq:bezier_delta_s,eq:bezier_zeta,eq:intensity_layer,eq:lambda_components} are matrices that are evaluated at each point on the wavenumber grid, for each value of $\mu$ in all layers of the atmosphere; for readability, the indices $\mu,\nu$ have been dropped, however.
While previous implementations have calculated such coefficients exclusively at the line-centre frequencies of each transition, the code presented here computes values for these coefficients at all points on the frequency grid.
This is to facilitate the inclusion of emission and absorption by other atmospheric constituents that factor into the balance of the statistical equilibrium equations.
Accordingly, the memory required for a computation can become quite large for a large frequency range, but the consideration of additional atmospheric constituents should have negligible impact.

\subsection{Opacity sampling}

\begin{figure}
    \centering
    \includegraphics[width=1.0\linewidth]{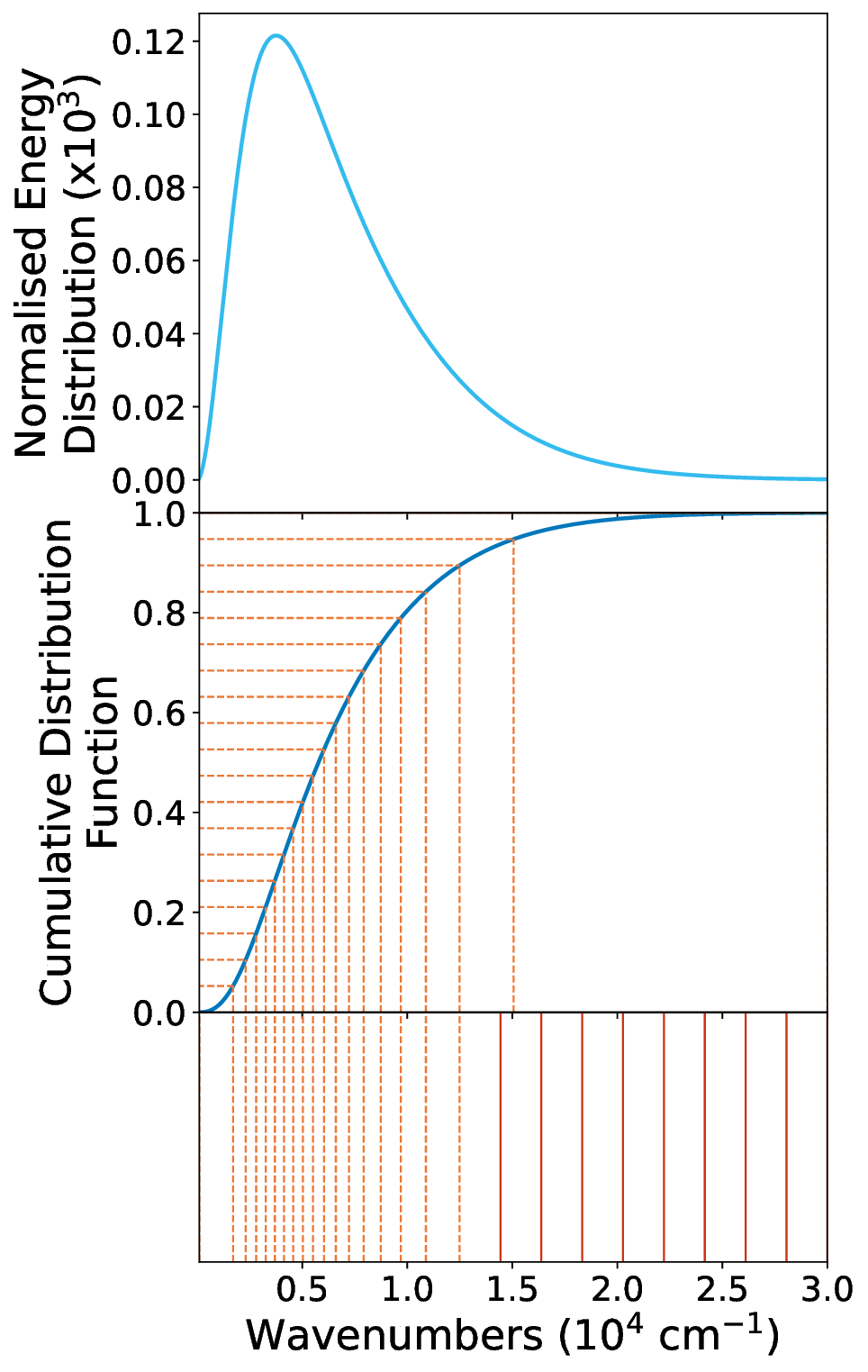}
    \caption{Example opacity sampling for the model KELT-20\,b atmosphere. The top panel shows the normalised energy distribution function across all layers of the atmosphere. The middle panel shows the cumulative distribution function (CDF) and the wavenumber positions corresponding to 20 samples evenly distributed in the CDF. The bottom panel shows the position of the final 27 grid points obtained, where the final 8 grid points marked with solid lines have been modified to ensure no grid spacing exceeds a maximal value, here 2\,000 cm$^{-1}$. The grid shown here is for illustrative purposes; models in this work utilise grids of 10\,000 points with a maximal spacing of 50 cm$^{-1}$.}
    \label{fig:opacity_sampling}
\end{figure}

As the B\'{e}zier interpolation coefficients, intensities and Lambda operator are computed for all layers of the atmosphere and at all angle and frequency points, it is pertinent to optimise the choice of grid points in order to avoid the allocation of extremely large arrays in memory.
To this end, the frequency grid points are chosen using an opacity sampling method \citep{98HeJo.nLTE} optimised for the temperatures structure of the model atmosphere.
Assuming initial LTE conditions, a cumulative distribution function is constructed as the sum of the mean intensity distributions at each layer, given by the Planck function.
The desired number of grid points are then picked corresponding to points linearly distributed through the cumulative distribution function, such that the energy density between each grid point is equivalent.
The grid points are chosen on a wavenumber scale as this is linear in terms of energy, unlike wavelength.
Modifications are made to the grid points chosen through this scheme to ensure that the wavenumber interval between successive points does not exceed a maximal value.
This is done to ensure that features arising from non-LTE effects that are not well described by the Planck distribution are not undersampled.
An example of this method applied for a small number of grid points is shown in \cref{fig:opacity_sampling}.
In the model presented here, the spectral grid was defined between 100--30\,000 cm$^{-1}$ for the atmospheric temperature structure described in Section \ref{sec:model_atmosphere} consisting of 10\,000 points.
A maximal spacing of 50 cm$^{-1}$ was applied which inserted an additional 99 points to ensure appropriate sampling.

\subsection{Vibrational aggregation and  line profiles}

A problem arises when trying to solve for a set of populations for molecules in non-LTE: even small diatomics have tens of thousands of energy levels, while larger polyatomics will have millions.
These in turn can give rise to billions of transitions even over a small frequency range; the ExoMol line list for CH$_{3}$Cl has over 166 billion transitions in the range 0--6400 cm$^{-1}$ \citep{jt733}.
Storing band profiles for billions of transitions in memory is computationally infeasible.
As such, the scale of the problem is reduced by allowing for energy levels to be aggregated on arbitrary characteristics and assigned aggregate IDs, such that transitions can be grouped together based on common upper and lower aggregate IDs, reducing the total number of effective transitions and hence profiles.
For diatomics, a straightforward aggregation is to group states together based on their vibronic assignment so that profiles are only computed for each vibronic band.
Doing so is analogous to treating the molecule as being in vibrational non-LTE, while leaving the rotations as LTE.
Hence, the vibronic band profiles are computed in the first iteration and stored to calculate the integrated mean intensity of the entire vibronic band at each iteration.
While this requires more memory to store potentially tens of thousands of vibronic band profiles for even simple diatomics, such as the OH radical for which the ExoMol line list contains 17\,822 distinct vibronic bands, the amount of time saved is considerable.
As only the normalised band profile is required in \cref{eq:u_ij,eq:v_ij}, rotational LTE is assumed when calculating the relative intensities of each rotational component of the vibronic band.

To ensure that the total relative intensity is conserved when computing the radiative transfer over an arbitrarily spaced grid, line profiles are computed as binned Voigt profiles treating the grid points as bin centres.
This makes use of the analytical solution to the integral of a Voigt profile obtained through solution to the integral using $N_{\rm GH}$ Gauss-Hermite quadrature points and weights, $t_{m}$ and $w_{m}$ \citep{jt708}, such that
\begin{equation}
    \label{eq:binned_gaussian}
    \begin{aligned}
        \bar{\sigma}^{fi}_{k} = \frac{I_{fi}}{\pi^{3/2}\Delta\tilde{\nu}}\sum^{N_{\rm GH}}_{m=1}w_{m}b_{m}\Big[&{\rm arctan(}z^{+}_{k,fi,m}{\rm)} \\
        & - {\rm arctan(}z^{-}_{k,fi,m}{\rm)}\Big], \\
    \end{aligned}
\end{equation}
where $\bar{\sigma}^{fi}_{k}$ is the integrated cross-section at the $k$-th bin, $\Delta\tilde{\nu}$ is the width of the bin, $z^{\pm}_{k,fi}$ is given by
\begin{equation}
    \label{eq:z_kfim}
    z^{\pm}_{k,fi,m} = \frac{\tilde{\nu}_{k} - \tilde{\nu}_{fi} - t_{m}\sigma_{G} \pm \frac{\Delta\tilde{\nu}^{\pm}}{2}}{\gamma_{L}},
\end{equation}
where $\sigma_{D}$ is the standard deviation of the Doppler profile, $\tilde{\nu}_{k}$ is the wavenumber of the bin centre, $\tilde{\nu}_{fi}$ is the transition wavenumber, $\gamma_{L}$ is the Lorentzian profile width.
The coefficient $I_{fi}$ is either the absorption or emission coefficient, which are given by \citet{jt708} and \citet{jt914}.
It is possible at this stage to apply a Doppler shift to the line centre to account for the effects of turbulent motion in each layer of the atmosphere.
$\Delta\tilde{\nu}^{+}$ is the wavenumber spacing between the next, higher-energy bin and $\Delta\tilde{\nu}^{-}$ is the spacing between the previous, lower-energy bin; $\Delta\tilde{\nu}$ is the sum of these values.
By considering these upper and lower widths separately, the integrated line profile can be computed over a grid with variable spacing.
The quantity $b_{m}$ is a correction factor given by
\begin{equation}
    \label{eq:b_m}
    b_{m} = \pi\left[{\rm arctan(}z_{{\rm end},fi,m}{\rm)} - {\rm arctan(}z_{{\rm start},fi,m}{\rm)}\right]^{-1},
\end{equation}
where
\begin{equation}
    \label{eq:z_startend}
    z_{{\rm start/end},fi,m} = \frac{\tilde{\nu}_{\rm start/end} - \tilde{\nu}_{fi} - t_{m}\sigma_{G}}{\gamma_{L}},
\end{equation}
and $\tilde{\nu}_{\rm start/end}$ refers to the starting and ending wavenumber points of the grid.
Computation of these profiles is highly parallelised and makes use of numerical optimisations made available through the \textsc{Numba} Python package \citep{Numba}.

\subsection{Radiative transfer solution}

The radiative transfer problem is solved first with an inward pass from the top of the atmosphere that propagates radiation towards the lower boundary of the atmosphere, updating the inward directed intensity and inward components of the Lambda operator following \cref{eq:intensity_layer,eq:lambda_layer}.
The source function and optical depths are then also updated for each layer.
The radiation field incident on the top of the atmosphere can be configured to an arbitrary value; see Section \ref{sec:model_atmosphere} for details on the configuration used in the models presented here.
Once the bottom of the atmosphere is reached, the radiation is propagated back up to the top of the atmosphere.
At the bottom of the atmosphere the initial outward directed radiation is taken to be that of a blackbody.
In the outward pass, both the inward and outward directed intensity and lambda operator are updated at each step.
For each layer that is not fixed to LTE, non-LTE level populations are obtained through solution to the statistical equilibrium equations given by \cref{eq:statistical_equilibrium}.
Given a matrix $\mathbf{Y}$ containing this set of linear equations, a vector containing the populations $\mathbf{n}$ is obtained via inversion of the expression
\begin{equation}
    \label{eq:matrix_equilibrium}
    \mathbf{Y}\mathbf{n} = \mathbf{R},
\end{equation}
where $\mathbf{R}$ is a column vector matching the first dimension of $\mathbf{Y}$.
It can be noted that the elements of the diagonal of $\mathbf{Y}$ contain the sum of the rates that $i$-th level corresponding to the row number is depopulated, while the off-diagonal elements are the rates at which that level is populated by the levels corresponding to the column indices.
In many implementations, \cref{eq:matrix_equilibrium} is solved by replacing one of the rows in $\mathbf{Y}$ with a conservation equation, setting all elements in the row to 1 and the correspond entry in $\mathbf{R}$ to 1.
Typically, this conservation equation is inserted as the last row and leaves $\mathbf{Y}$ as an invertible square matrix. 
This method ensures that the populations obtained from the solution sum to 1, and ensures $\mathbf{Y}$ is non-singular.
It can be seen that without the addition of the conservation equation, $\mathbf{Y}$ is singular as any linear multiple of $\mathbf{n}$ would also solve \cref{eq:matrix_equilibrium}.
As each radiative and collisional rate term in \cref{eq:statistical_equilibrium} appears in multiple rows, i.e.: the rows corresponding to the levels that the rate coefficient is describing the population or depopulation of, insertion of the conservation equation is valid as the system of equations is overdetermined.
However, when chemical or collisional creation or destruction rates which act only on a single level are included, this is not true.
Even without such rates, it was found that the population of the level corresponding to the index of the row where the conservation equation was inserted was biased towards larger values.
While these effects could be on the order of $<10^{-10}$, this is impactful for highly excited states where this is on the same order as the population. 
To avoid this, \cref{eq:matrix_equilibrium} is solved using an overdetermined rectangular form of $\mathbf{Y}$ where the conservation equation is added as an additional row, instead of overwriting another.
By computing a Moore-Penrose pseudo-inverse using singular-value decomposition, populations are obtained that give a root mean square error consistently 3 orders of magnitude smaller than those obtained through direct inversion of a square $\mathbf{Y}$ matrix.

\subsection{Populations}

The statistical equilibrium is solved for the aggregated vibronic level populations, meaning the relative excitation of the rotational, parity, fine structure, etc., sub-levels of the vibronic states are assumed to be in LTE.
Hence, populations are obtained for all sub-levels in non-LTE, $n_{i}$, from the non-LTE vibronic population obtained through solution to \cref{eq:matrix_equilibrium}, $\tilde{n}_{k}$, with the relationship 
\begin{equation}
    \label{eq:scaled_populations}
    n_{i} = \tilde{n}_{k} \frac{n_{i}^{*}}{\tilde{n}_{k}^{*}},
\end{equation}
where $n_{i}^{*}$ is the Boltzmann population of the sub-level, $\tilde{n}_{k}^{*}$ is the aggregated Boltzmann population of the vibronic level and the $i$-th sub-level is a member of the $k$-th vibronic level.
The Boltzmann populations are calculated from the equation
\begin{equation}
    \label{eq:boltzmann_pop}
    n_{i}^{*} = \frac{g_{i}\,\exp\left(-hc\tilde{E}_{i}/k_{\rm B}T\right)}{Q\left(T\right)},
\end{equation}
where $Q\left(T\right)$ is the partition function given by
\begin{equation}
    \label{eq:partition_func}
    Q\left(T\right) = \sum_{i} g_{i}\,\exp\left(-hc\tilde{E}_{i}/k_{\rm B}T\right).
\end{equation}
Under this formulation, different vibronic bands are effectively represented by different temperatures, while the rotational motion within the band is thermally equilibrated.

\subsection{Successive over-relaxation \& damping}

A successive over-relaxation (SOR) scheme is employed to increase the rate of convergence, based on the implementation outlined by \citet{95TrFaxx.nLTE}.
The vector of populations obtained at each iteration $\mathbf{n}$ from \cref{eq:matrix_equilibrium} is compared to the populations from the previous iteration, $\mathbf{n}^{\dagger}$ such that
\begin{equation}
    \label{eq:population_change}
    \mathbf{n} = \mathbf{n}^{\dagger} + \delta\mathbf{n},
\end{equation}
where $\delta\mathbf{n}$ is a vector describing the change in each level population.
In a successive over-relaxation scheme, the change in populations in each iteration is modified by the relaxation parameter $0 \leq \omega \leq 2$ such that
\begin{equation}
    \label{eq:sor_population_change}
    \mathbf{n} = \mathbf{n}^{\dagger} + \omega\delta\mathbf{n}.
\end{equation}
An optimum increase in the rate of convergence can be achieved in a Gauss-Seidel iteration by taking the relaxation parameter to be
\begin{equation}
    \label{eq:_sor_omega}
    \omega = \frac{2}{1 + \sqrt{1 - \rho}},
\end{equation}
where $\rho$ is the spectral radius of the iteration matrix \citep{71Young.book}.
\citet{95TrFaxx.nLTE} pointed out that a convenient approximation of $\rho$ can be obtained by considering the convergence rate over the last two iterations,
\begin{equation}
    \label{eq:rho}
    \rho \approx \frac{{\rm max}\left(|\mathbf{n} - \mathbf{n}^{\dagger}|/\mathbf{n}^{\dagger}\right)}{{\rm max}\left(|\mathbf{n}^{\dagger} - \mathbf{n}^{\dagger\dagger}|/\mathbf{n}^{\dagger\dagger}\right)},
\end{equation}
where $\mathbf{n}^{\dagger\dagger}$ is the vector of populations obtained two iterations prior and $\max\left(|\cdot|\right)$ denotes the largest absolute element in the vector or $l^{\infty}$ norm.
Given this formulation requires two previous iterations to compare populations, the SOR step cannot be enabled immediately.
Indeed, the SOR correction is only enabled when the largest relative change in populations across all layers of the atmosphere satisfies the criterion ${\rm max}\left(|\mathbf{n} - \mathbf{n}^{\dagger}|\right) \leq 0.1$ at the end of an iteration.
As the value of $\omega$ is recalculated at each iteration, the iteration scheme is nonstationary.
Though setting the relaxation parameter this way should ensure the values of $\mathbf{n}$ stay within physically meaningful bounds, checks are made to ensure that no elements of $\mathbf{n}$ are over-corrected to negative values.

In some situations, the iterations can get stuck between two solutions and oscillate between them.
In these cases, the population changes between iteration are too large and the model is likely overshooting the stable minima.
This can be detected by identifying when the product of the differences between sequential maximum relative population changes has a negative sign, i.e.: the maximum relative change repeatedly gets larger and then smaller.
When this is detected, successive over-relaxation is disabled and instead damping is turned on.
Damping is performed by setting the updated populations to the summation of the new and previous populations, weighted by a damping factor $\epsilon$ such that
\begin{equation}
    \label{eq:damping}
    \mathbf{n}^{\rm damp.} = \epsilon\mathbf{n} + \left(1 - \epsilon\right)\mathbf{n}^{\dagger}.
\end{equation}
Here the damping parameter was set to $\epsilon = 0.1$ though this value is configurable.
In cases where damping is enabled, the convergence criteria is relaxed from a maximum relative change of $10^{-3}$ to $5\times10^{-3}$.
Improvements to the stability of cases where damping is required will be addressed in future releases, and methods may be implemented to set $\epsilon$ dynamically, as is done for the successive over-relaxation parameter.

\section{Collisional data}

Collisions are the process that ensure the populations of an atom or molecule are coupled to the thermal motion of the gas.
This in turn means that they are the key process for driving a gas into LTE and as such LTE populations should be obtained when solving the statistical equilibrium equations in the denser layers of an atmosphere.
Unfortunately, the provision of accurate collisional rate data covering a comprehensive range of temperatures is extremely limited.
For OH, much of the collisional rate data available in the literature focuses on reactive and non-reactive collisions with O, O$_{2}$ and N$_{2}$, as these species are important for modelling observed OH emission in Earth's upper atmosphere \citep{89ReJeCr.OH,90SaCo.OH,91DoLiBl.OH,93ChCo.OH,96KnDyCo.OH,97DyKnCo.OH,97Adler.OH,03LaDyCo.OH,04Varandas.OH,10voHaFu.nLTE,12XuGaSm.OH,13CaHoVa.OH}.
Chemical formation rates for the O$_{3}$ + H reaction are also important there, as the reaction preferentially forms OH in the $v = 9$ state \citep{97Adler.OH}.
Most of these data are limited to temperatures applicable to Earth's atmosphere, considerably below the temperatures of hot and ultra-hot Jupiters.
However, these species are not considered in the model atmosphere discussed in Section \ref{sec:model_atmosphere}, where the dominant atmospheric constituents are H$_{2}$, H and He.
As the collisional rates scale linearly with number density, collisional transitions with these species will dominate compared to those with any other.
The rate data employed in the models presented here are summarised in \cref{tab:oh_collision_rates} and discussed below.

\citet{13KoYaKa.OH} measured the vibrational quenching rates of OH $\left(v^{\prime} = 1-4\right)$ by Helium and used values for higher vibrational states measured by \citet{90SaCo.OH} and \citet{97DyKnCo.OH} to fit an analytical expression for the quenching rate up to $v = 12$.
These rates are single-quantum, meaning the vibrational state of the OH molecule undergoes $\Delta{}v=v^{\prime}-v^{\prime\prime}=1$; collisional rates with $\Delta{}v>1$ are assumed to be negligible.

\citet{06AtAl.OH} calculated vibrational quenching rates for OH $\left(v^{\prime} = 1-2\right)$ by atomic Hydrogen, observing that they exhibited multi-quantum behaviour.
Indeed, the $v = 2 \rightarrow 0$ rate was measured to be roughly twice as fast as the $v = 2 \rightarrow 1$ rate.
OH collisions with O and O$_{2}$ are also known to exhibit multi-quantum behaviour \citep{97Adler.OH,04Varandas.OH}.
The collisional rates with atomic Hydrogen are between three and seven orders of magnitude greater than those for Helium, with the largest discrepancy occurring at low $v$.
This means the layer of the atmosphere at which LTE behaviour is recovered will be primarily determined by the collision rates with Hydrogen.
Accordingly, vibrational quenching rates arising from collisions with the other atmospheric constituents are neglected, either due to low rates or low abundance.
Indeed, vibrational quenching rates arising from collisions with H$_{2}$O are known to be small \citep{06McRaMa.OH}.
For collisions with CO, the vibrational quenching rate is comparable to that of H \citep{02BiMuYu.OH} but the absolute rate will be significantly smaller due to the lesser CO abundance.

The choice of the molecular Hydrogen dissociation pressure has a considerable impact on pressure at which OH transitions from LTE to non-LTE.
This is because, similar to Helium, the vibrational quenching rates of OH by molecular Hydrogen are three to four orders of magnitude smaller than those of atomic Hydrogen, \citep{76StJo.OH,82GlEnCh.OH,93ChCo.OH,11BaPaMe.OH}.
The H$_{2}$ collisional rates for $v^{\prime} = 4 - 9$ are taken from \citet{76StJo.OH} as they provide data for the largest set of OH vibrational states; the majority of other publications only provides upper bounds on the OH + H$_{2}$ quenching rate.
Collisional rates are extrapolated down to $v^{\prime} = 1$ and up to $v^{\prime} = 10$ assuming a conservative exponential relationship, consistent with the reported values.
The collisions are also assumed to be single-quantum.

It is possible that in regions where H$_{2}$ is the dominant atmospheric species the resultant collisions will be insufficient to drive OH into LTE.
Once the H$_{2}$ begins to dissociate, however, the OH level populations may revert back to LTE.
It can be seen from the available rate data that even if 0.01\% of the H$_{2}$ has dissociated, the overall vibrational quenching rate of the OH $v=1$ level will be greater due to expedient quenching by atomic Hydrogen.

Vibrational quenching rates for OH + H collisions were extrapolated up to $v^{\prime} = 10$, assuming single-quantum behaviour due to a lack of measurements to quantify the multi-quantum behaviour.
Extrapolation was done using an exponential energy difference relation, similar to that used by \citet{13KoYaKa.OH},
\begin{equation}
    \label{eq:rates_approx}
    k_{v \rightarrow v - 1} = \exp\left(a - b\Delta{}E\right),
\end{equation}
where $\Delta{}E$ is the energy difference between the $v$ and $v - 1$ vibrational levels and $a$ and $b$ are constants which were fit to $-21.6$ and $0.000266$, respectively.
Collisional rates approximated this way were rounded to 2 significant figures and are highly uncertain.
This fit represents a conservative extrapolation, where the OH $\left(10 \rightarrow 9\right)$ quenching rate is only 50\% larger than the OH $\left(10 \rightarrow 9\right)$ rate.
To assess the impact of these rate data on the obtained non-LTE populations, collisional rates were fit in conjunction with the radiative rates to obtained rates sufficient to recover LTE populations.
These rates, which were intended to enable a ``smooth'' transition between LTE and non-LTE are given in \cref{tab:oh_collision_rates}.
The extrapolated rates are highly uncertain, though models using the larger OH + H quenching rates recover LTE behaviour well at higher pressures, suggesting they are of appropriate relative magnitudes.

Once downward collisional rates are obtained, the upward rates are obtained through the detailed balance relationship such that
\begin{equation}
    \label{eq:collision_detailed_balance}
    C_{ji} = C_{ij} \frac{g_{i}}{g_{j}} \exp\left(-\frac{h\nu_{ij}}{k_{\rm B}T}\right),
\end{equation}
in the case that $E_{i} > E_{j}$.
The effect of the detailed balance of upward and downward collisional rates then recovers the Boltzmann distribution of the level populations and ensures that a result closely matching LTE conditions is obtained.

\begin{table*}
    \centering
    \begin{tabular}{ccccc}
        \hline
         & \multicolumn{2}{c}{OH + H} & OH + H$_{2}$ & OH + He \\
        Collision ($v^{\prime} \rightarrow v^{\prime\prime}$) & Conservative Fit & Smooth Fit & & \\
        \hline
        $1 \rightarrow 0$ & \multicolumn{2}{c}{1.600$\times10^{-10a}$} & 1.0$\times10^{-14}$ & 3.2$\times10^{-17c}$ \\
        $2 \rightarrow 0$ & \multicolumn{2}{c}{1.043$\times10^{-10a}$} & - & - \\
        $2 \rightarrow 1$ & \multicolumn{2}{c}{0.654$\times10^{-10a}$} & 4.0$\times10^{-14}$ & 1.4$\times10^{-16c}$ \\
        $3 \rightarrow 2$ & 1.8$\times10^{-10}$ & 5.8$\times10^{-10}$ & 9.0$\times10^{-14}$ & 4.4$\times10^{-16c}$ \\
        $4 \rightarrow 3$ & 1.8$\times10^{-10}$ & 1.5$\times10^{-9}$ & 1.8$\times10^{-13b}$ & 1.2$\times10^{-15c}$ \\
        $5 \rightarrow 4$ & 1.9$\times10^{-10}$ & 4.0$\times10^{-9}$ & 3.9$\times10^{-13b}$ & 3.2$\times10^{-15e}$ \\
        $6 \rightarrow 5$ & 2.0$\times10^{-10}$ & 9.8$\times10^{-9}$ & 6.8$\times10^{-13b}$ & 8.2$\times10^{-15e}$ \\
        $7 \rightarrow 6$ & 2.1$\times10^{-10}$ & 2.4$\times10^{-8}$ & 8.0$\times10^{-13b}$ & 2.1$\times10^{-14e}$ \\
        $8 \rightarrow 7$ & 2.2$\times10^{-10}$ & 7.6$\times10^{-8}$ & 7.6$\times10^{-13b}$ & 5.1$\times10^{-14e}$ \\
        $9 \rightarrow 8$ & 2.3$\times10^{-10}$ & 1.8$\times10^{-7}$ & 5.8$\times10^{-13b}$ & 1.3$\times10^{-13e}$ \\
        $10 \rightarrow 9$ & 2.4$\times10^{-10}$ & 5.1$\times10^{-7}$ & 4.4$\times10^{-13}$ & 3.4$\times10^{-13d}$ \\
        \hline
    \end{tabular}
    \caption{Collision rates in cm$^{3}$s$^{-1}$ for vibrational quenching of OH($v^{\prime} \rightarrow v^{\prime\prime}$) by atomic Hydrogen, H$_{2}$ and atomic Helium. Values marked with $^{a}$ are calculated by \citet{06AtAl.OH}; $^{b}$ were determined from experiment by \citet{76StJo.OH}. For the OH + He rates, those marked with $^{c}$ were determined from experiment by \citet{13KoYaKa.OH} and the value marked with $^{d}$ was measured by \citet{97DyKnCo.OH}; those marked with $^{e}$ were determined from a fit to the available experimental data by \citet{13KoYaKa.OH}.}
    \label{tab:oh_collision_rates}
\end{table*}

Data for all of the collisions considered here are extremely limited in the temperature ranges they cover.
While some collisional rates for OH show an increase over some temperature ranges, this varies depending on the collisional partner.
\citet{24ZaTaHa.OH} extrapolated the OH + H collisional rates of \citet{06AtAl.OH} to higher temperatures, assuming rates at 1000 K that were 3 orders of magnitude smaller than those at calculated for 300 K.
As the effects of temperature are not well understood, values are taken from the literature as is and no attempt is made to extrapolate the rates to higher temperatures.
It should be noted that if the collision rates are smaller at these higher temperatures, then the number density and hence pressure threshold required for collisions to drive OH into LTE will be higher.
In turn this will mean that OH will be driven into non-LTE at lower altitudes in the atmosphere, likely further increasing the impact of non-LTE effects on observed spectra.

Limited rates for vibrational, electronic, or vibronic collisional quenching of OH are available for the first two vibrational levels of the excited A~$^{2}\Sigma^{+}$ state \citep{95Paul.OH,17MaGaDi.OH}.
These rates are generally provided as total quenching rates, however, with no specified lower state.
To avoid further assumptions, the model was limited to the levels up to $v = 10$ in the OH X~$^{2}\Pi$ ground state.
Model spectra were also limited to transition energies below 30\,000 cm$^{-1}$, below the first A~$^{2}\Sigma^{+}$-X~$^{2}\Pi$ fundamental band.

\section{Model atmosphere}
\label{sec:model_atmosphere}

\begin{figure*}
    \centering
    \includegraphics[width=\linewidth]{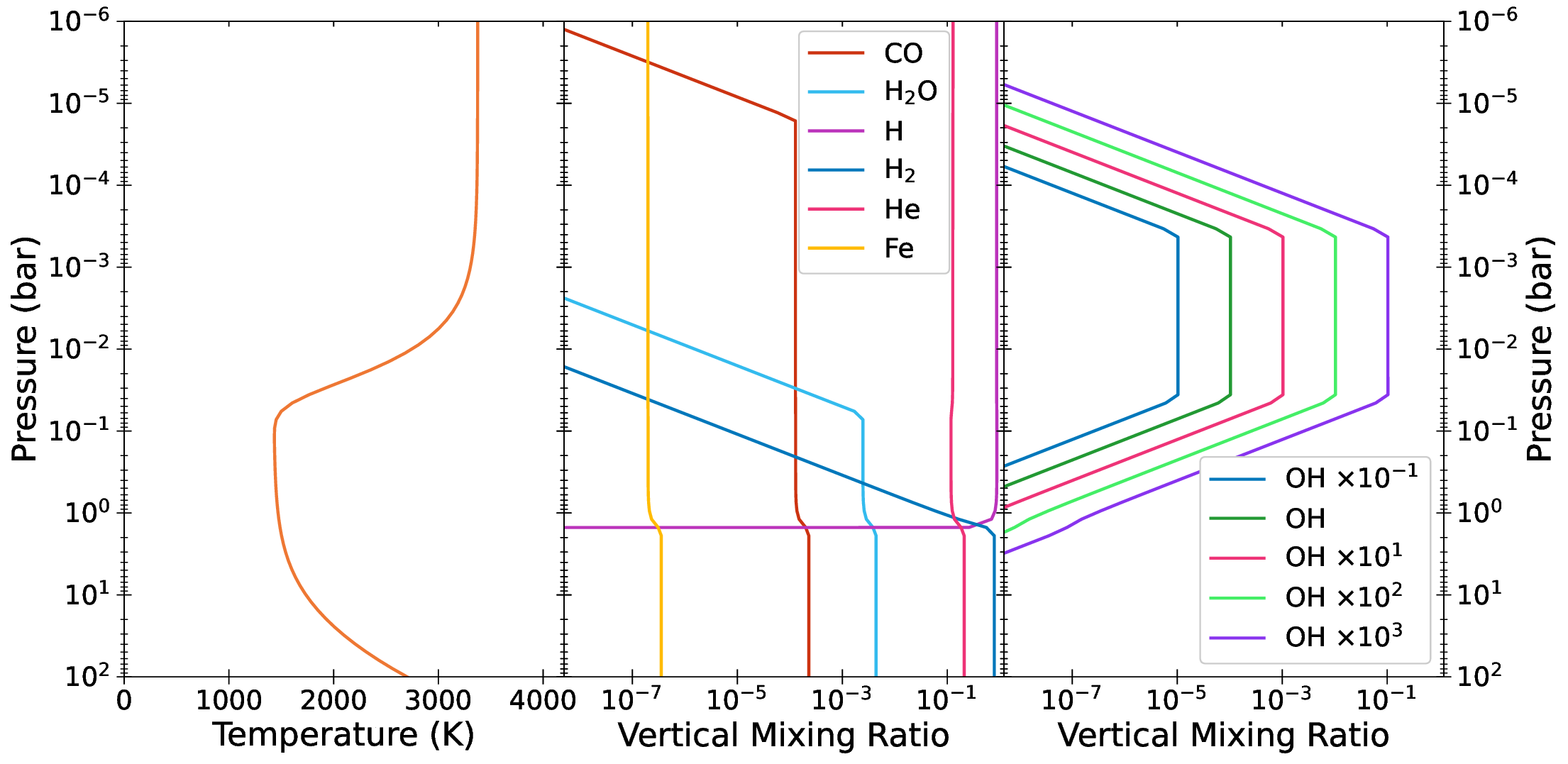}
    \caption{Leftmost plot shows the temperature of the atmosphere as a function of the pressure; lower pressures correspond to altitudes higher in the atmosphere. Centre plot shows the vertical mixing ratios for the atmospheric constituents excluding OH. Rightmost plot shows the five test case vertical mixing ratio scenarios for OH. Profiles are adapted from \citet{25FiXiXu.exo} except for H and H$_{2}$ which are based on their retrieved total Hydrogen mass fraction. The dip in the vertical mixing ratios of Fe, CO, H$_{2}$O and He around 1 bar is due to the significant change in mean molecular weight from the dissociation of H$_{2}$.}
    \label{fig:kelt-20b_profiles}
\end{figure*}

The ultra-hot Jupiter KELT-20\,b (also MASCARA-2\,b) is used as a case study to illustrate the functionality of the TIRAMISU code and also to test the wavelengths at which non-LTE effects may be detectable.
\citet{25FiXiXu.exo} recently performed an analysis of the ultra-hot Jupiter KELT-20\,b, looking at the dissociation of atmospheric H$_{2}$O leading to the production of OH.
OH produced in this manner is believed to be important in the atmospheres of several exoplanets \citep{24GaLaSn.H2O,25LoBeSi.exo,25FiXiXu.exo} and may be produced with a highly non-LTE population distribution.
The retrieved median temperature/pressure profiles are taken to construct a model atmosphere for assessing the impacts of non-LTE effects in the spectra of OH.
Mass mixing ratios for H$_{2}$O, CO, OH and Fe are taken from their retrieved model, as well as a fixed Hydrogen abundance.
Opacities are obtained from the ExoMol database for H$_{2}$O \citep{jt734}, CO \citep{15LiGoRo.CO} and OH \citep{jt969}, including OH photoabsorption data \citep{jt982}.
The Fe opacities were calculated from the Kurucz database \citep{11Kurucz.db}.
The retrieved molecular Hydrogen dissociation pressure was poorly constrained so is fixed here to 1 bar for simplicity; the Hydrogen mass fraction at lower pressures is assumed to be comprised of atomic Hydrogen exclusively.
The OH mass mixing ratio is varied to test the sensitivity of observable non-LTE features to the OH abundance; in all cases, the remaining mass of the atmosphere is assumed to be Helium.
Mass mixing ratios are converted to vertical mixing ratios as described by the authors.
These profiles are shown in \cref{fig:kelt-20b_profiles} and divide the atmosphere into 80 layers with even logarithmic pressure spacing ranging from $10^{2} - 10^{-6}$ bar.

KELT-20\,b orbits an A2V host star, the radiation from which impacts the radiative transfer within the planet's atmosphere.
KELT-20 is configured to emit as a blackbody with $T_{\rm eff} = 8730$ K and this radiation is incident on KELT-20\,b orbiting at a distance of 0.057 AU and having a planetary radius of $R_{p} = 1.83 R_{J}$.
These physical parameters of KELT-20\,b are taken from the work of \citet{18TaJuAl.exo} and the stellar parameters from \citet{17LuRoZh.exo}.

The lower atmosphere is likely to be in LTE, but the exact boundary is hard to determine.
To simplify the model, the first 20 layers of the atmosphere with pressures ranging from $10^{2} - 1$ bar are fixed to LTE, coinciding with the assumed H$_{2}$ dissociation pressure boundary.
The 60 layers above those are floated, meaning populations are computed via the Gauss-Seidel scheme outlined in Section \ref{sec:method}.

\section{Results}

\begin{figure*}
    \centering
    \includegraphics[width=\linewidth]{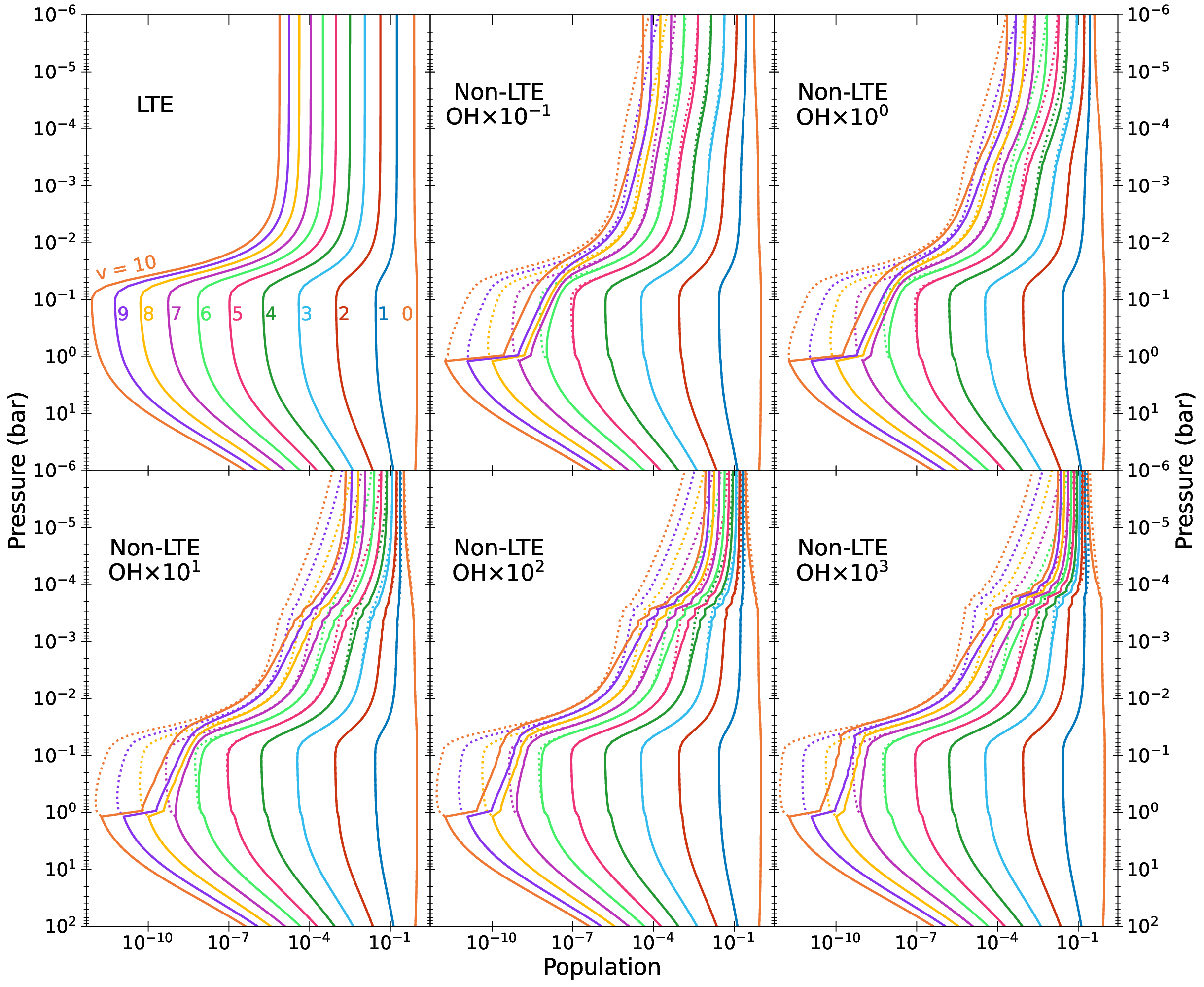}
    \caption{Variation in the populations of the $v = 0 - 10$ vibrational levels of the OH X~$^{2}\Pi$ ground state as a function of pressure in the model KELT-20\,b atmosphere. The left profile shows the population of $v = 10$, decreasing to $v = 0$ for the rightmost profile. Populations shown above 1 bar are obtained from solution to the statistical equilibrium equations; populations below 1 bar are fixed to LTE and obtained from the Boltzmann distribution. The top left panel shows the population profiles for an atmosphere entirely in LTE and the five other plots show the non-LTE scenario with varying OH abundances. For the non-LTE models, solid lines represent populations obtained using the conservative fit to the OH + H collisional rate data and dotted lines represent populations obtained using the fit to the rate data intended to smooth the transition between LTE and non-LTE.}
    \label{fig:population_profiles}
\end{figure*}

\begin{figure}
    \centering
    \includegraphics[width=\linewidth]{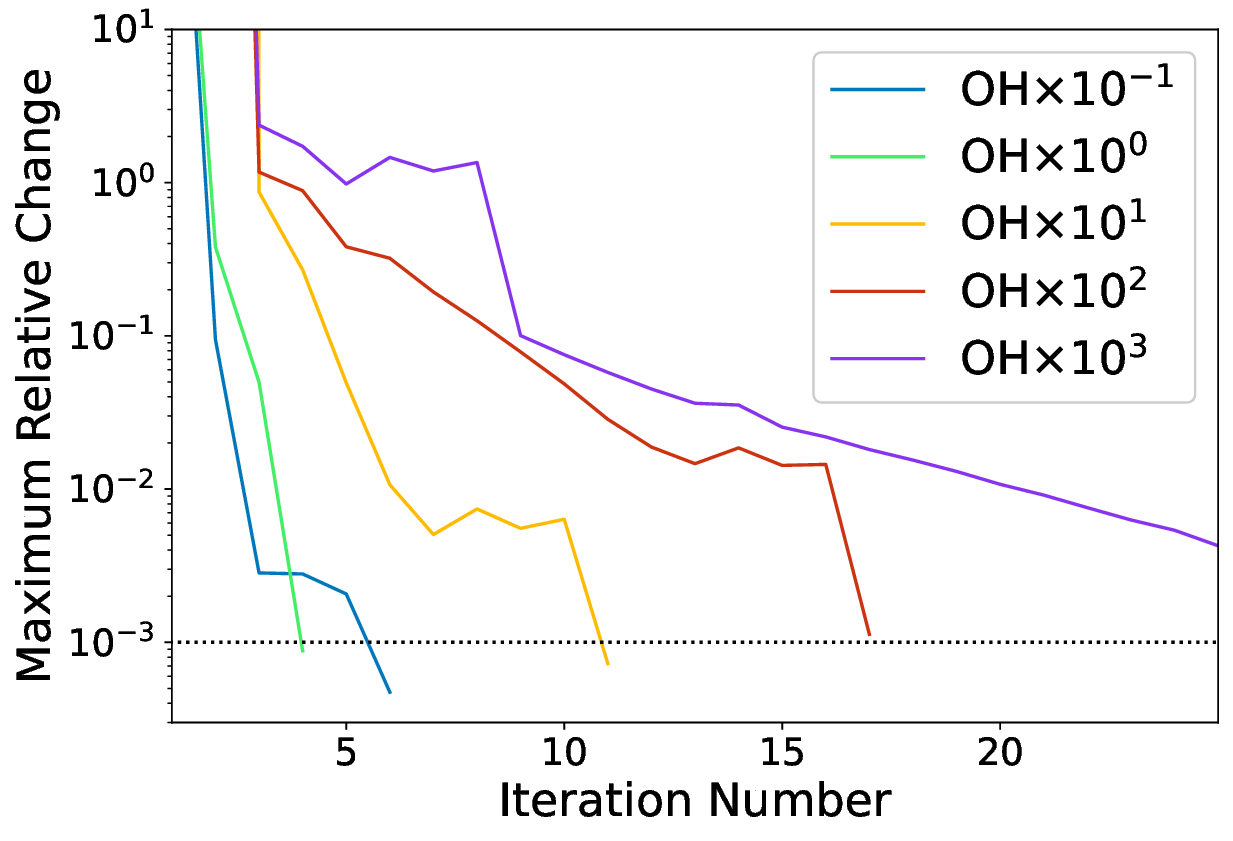}
    \caption{Convergence behaviour of OH in non-LTE with varying abundance scaling. The maximum relative change of a population throughout the atmosphere is shown for each iteration. The convergence threshold of a maximum relative change of $10^{-3}$ is marked with a horizontal dotted black line.}
    \label{fig:convergence}
\end{figure}

The non-LTE radiative transfer problem is solved for the model KELT-20\,b atmosphere and the converged populations are shown as a function of pressure in \cref{fig:population_profiles}.
Models were run for five OH vertical mixing ratio scenarios, where the OH profile retrieved by \citep{25FiXiXu.exo} was scaled by 10$^{x}$ with $x$ taking integer values from -1 to 3.
The OH profiles are constrained by the H$_{2}$O dissociation pressure in the lower atmosphere which drives the production of OH in the layer at 3.6$\times10^{-2}$ bar and above.
OH abundance decrease above the assumed dissociation pressure, affecting the layers of the atmosphere with a pressure of 3.4$\times10^{-4}$ bar and below.
In all scenarios tested, the OH abundance is roughly constant in the region bounded by these two dissociation pressures; outside this region the abundance exhibits a quartic decay as a function of pressure.

In the case of the conservative fit to the OH + H vibrational quenching rates, the switch between LTE and non-LTE populations in the five models is relatively smooth for the $v =  0 - 6$ levels, though there are discontinuities for the $v = 7 - 10$ levels.
This is indicative of inaccuracies in the vibrational quenching rates for these more highly-excited levels, which are not large enough to drive these populations into LTE.
It can be seen that as the OH abundance increases, the magnitude of this discontinuity decreases, with the switch between LTE and non-LTE being smooth for the $v = 7$ level in the three models with the most OH.
In the models run with the ``smooth'' fit to the collisional rate data, the transitions from LTE to non-LTE are much smoother.

As the OH abundance increases, the degree to which the populations deviate from LTE also increases.
In the OH$\times10^{-1}$ model, 53.4\% of the population is found to be in the ground vibronic state in non-LTE at the top of the atmosphere, compared to 76.7\% in LTE.
As the abundance increases up to the OH$\times10^{3}$ model, this proportion decreases down to only 23.1\% of the population being in the vibronic ground state.
While the total OH vertical mixing ratio is small at the top of the atmosphere, in the conservative rates scenarios there is significantly enhanced vibrational excitation compared to LTE from roughly 2$\times10^{-3}$ bar and above.
In the smooth rates scenarios, this meaningful deviation from LTE begins higher in the atmosphere at roughly 8$\times10^{-4}$ bar.
In both cases, the quantities of OH in these regions are sufficient to have an observable impact on the emission spectra; see Section \ref{subsec:observable_features} for examples.

The population profiles exhibit several noteworthy features: first, there is a ``bump'' in the region that the temperature profile transitions from a temperature inversion to linear, around $10^{-2}$ bar.
This also coincides with the region where OH begins to form.
Above this, the gradient of the population profiles decrease until the region that the OH begins to dissociate.
Here, another bump is observed as the more highly excited states begin to become rapidly more populated.
Above the OH dissociation region, the population profiles tend towards linear as they are only coupled to the linear temperature profile.

The scenarios with the three lowest OH abundances in \cref{fig:population_profiles} all show deviation from LTE in the region between 1.0 and 3.6$\times10^{-2}$ bar in the 3 vibrational levels with $v = 8 - 10$.
This is likely due to the OH abundances in these scenarios being too small in this region for variations in the populations of these levels to have a meaningful impact on the overall opacity.
Reinforcing this, the populations for these levels in the scenarios with 10$^{2}$ and 10$^{3}$ times the initial OH abundance do begin to recover the LTE behaviour at these pressures.
In the two highest abundance scenarios, the level populations deviate less from the LTE case until pressures below the OH dissociation pressure which can be seen as a bump in the population profiles.

The number of iterations and hence time taken to converge increases as a function of the OH abundance.
This is an expected behaviour, as the fraction of the intensity which is determined by non-LTE effects and hence changes each iteration is also increasing with the OH abundance.
The convergence behaviour is shown in \cref{fig:convergence}.
As expected, the models with the smallest OH abundance converge extremely quickly as the overall fraction of the atmosphere in non-LTE is extremely small.
The two models with the lowest OH abundances converge in roughly 1 hour, increasing up to roughly 5 hours for the models with the highest abundances.
With the current implementation, the model takes approximately 10 minutes per iteration, primarily owing to the sequential evaluation of each layer.
The is necessary due to the construction of the terms in \cref{eq:intensity_bezier,eq:lambda_layer} which require updated coefficients for the previous layers.
While these could be reformulated to use coefficients exclusively from the previous iteration as in MALI, the overall number of iterations required to converge should then increase.

\subsection{Population inversions \& masing}

\begin{figure*}
    \centering
    \includegraphics[width=1.0\linewidth]{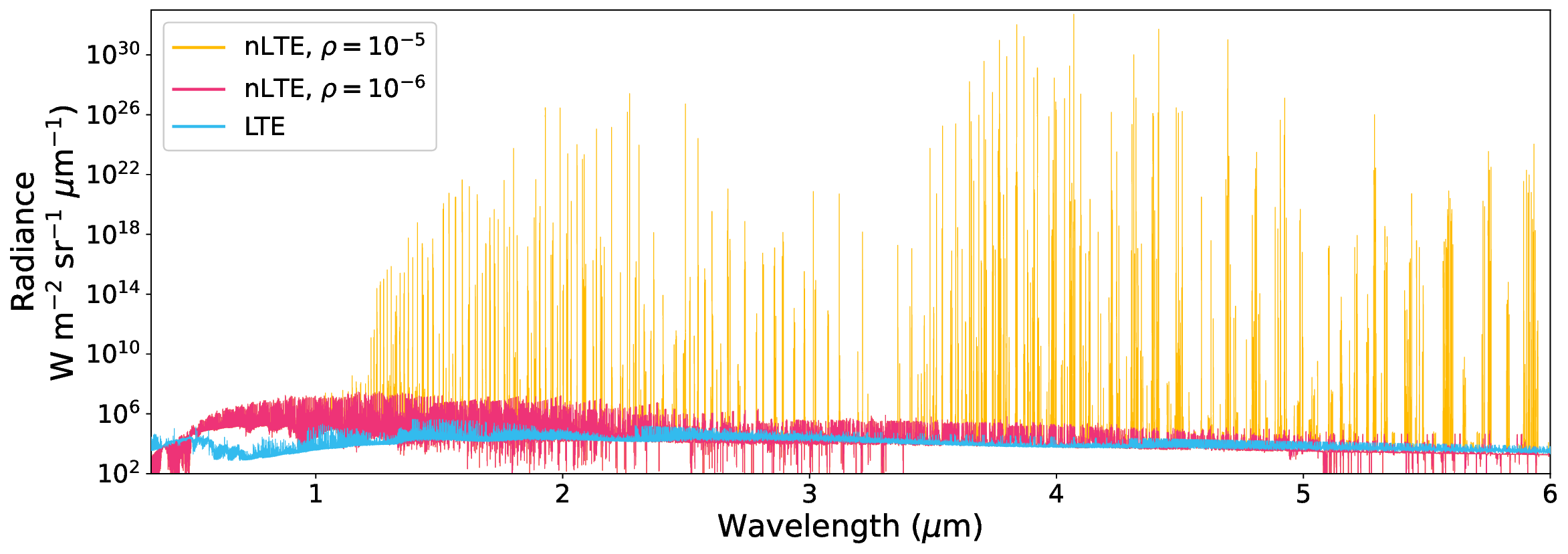}
    \caption{Sensitivity of intensity calculations on the masing-damping parameter $\rho$ in the presence of population inversions. The emergent emission at the top of the atmosphere under LTE is shown, in contrast to two non-LTE scenarios with different values of $\rho$. The masing transitions arising from the $\rho = 10^{-5}$ model are $10^{25}$ times more intense than those from the $\rho = 10^{-6}$ model. Models using $\rho < 10^{-6}$ are indistinguishable from those with $\rho = 10^{-6}$, indicating masing is effectively quashed in this regime. When $\rho$ is increased up to $10^{-5}$ however, the masing intensities are unphysical and the result of numerical overflows.}
    \label{fig:masing_test}
\end{figure*}

It can be seen from \cref{eq:chi_ij} that the absorption coefficient is not rigorously positive.
Indeed, in the case of a population inversion where the upper state of a transition has a greater population, this term can be negative.
Under LTE conditions this cannot occur as the Boltzmann distribution strictly decreases as a function of energy.
In non-LTE however, this simply leads to the dominance of stimulated emission over absorption and is commonly observed as the cause of masers in a variety of astronomical media.
A complication arises however, as a negative absorption cross section produces negative opacities, which the interpolation coefficients used in short characteristics methods are not generally designed for.
It can be seen from \cref{eq:bezier_coefs} that the coefficients are defined for negative values of $\delta^{\pm}_{l}$, but it requires careful limiting of the control points given in \cref{eq:bezier_control_point,eq:bezier_c0_c1} to maintain the desired numerical behaviour, namely positive intensities.
Further development is needed to ensure these calculations are resilient to negative opacities and will be discussed in a later work.
To circumvent this challenge for now, an effective absorption coefficient is considered as in \citet{97YaFiGr.nLTE} such that
\begin{equation}
    \label{eq:chi_cap}    
    \chi_{\mu,\nu,l}^{\rm eff.} = \begin{cases}
    \xi\,max\left(\chi_{\mu,\nu,l}\right)\exp{\left(-\lvert\chi_{\mu,\nu,l}\rvert\right)} & \hfill\textrm{if } \chi_{\mu,\nu,l} < 0\\
    \chi_{\mu,\nu,l} & \hfill\textrm{otherwise.}
    \end{cases}
\end{equation}
Here a value of $\xi = 0.1$ is adopted, similar to the method employed by \citet{15Nesterenok.nLTE}. 
This effective absorption cross section is used in the calculation of an effective source function and optical depths which are then used in the construction of the B\'{e}zier interpolants in \cref{eq:bezier_coefs}.
The real absorption cross section is used in the formal solution, however.

Given the optical depth is dependent on the absorption cross section, the number density of the species in question and the path length through the medium, it is very easy for the terms in the formal solution which depend of $\exp{\left(-\tau\right)}$ to become very large.
This is particularly true in model atmospheres using sparse pressure grids, such that each layer of the atmosphere corresponds to a very large path length.
Indeed, in the model considered here the upper layers of the atmosphere can have a thickness on the order of $10^{7}$ m.
Accordingly, calculation of the emergent intensity through thick or dense layers that are determined to have population inversions can result in catastrophic numerical overflows, leading to the determination of infinite intensities at the top of the atmosphere.
To rectify this, the negative components of the opacity and source function must also be capped in the calculation of the formal solution to avoid unphysical behaviour.
The negative components of $\chi_{\mu,\nu,l}$ are capped at $-\rho\,{\rm max}\left(\chi_{\mu,\nu,l}\right)$.
While this maintains the negative opacity and hence masing behaviour, it was determined that for the non-LTE model atmosphere presented here a value of $\rho<10^{-6}$ was necessary to avoid numerical overflows.
If $\rho$ was set to $10^{-5}$, masing resulted in emergent intensities in excess of $10^{32}$ W m$^{-2}$ sr$^{-1}$ ${\mu}$m$^{-1}$, whereas if $\rho$ was set to $10^{-6}$ or smaller, the intensity of masing transitions did not exceed the maximum intensity of the rest of the emergent emission, which was on the order of $10^{7}$ W m$^{-2}$ sr$^{-1}$ ${\mu}$m$^{-1}$.
The emergent intensity in this non-LTE case is much larger than the LTE case for the same physical setup, which had a maximum emergent intensity on the order of $10^{5}$ W m$^{-2}$ sr$^{-1}$ ${\mu}$m$^{-1}$.
These emission spectra are shown in \cref{fig:masing_test}. 
This is due to the masing test setup having an extreme population inversion just below the H$_{2}$ dissociation threshold and in the top half of the atmosphere where the H number density is low, allowing for significant intensity to accumulate throughout the atmosphere without significant opacity to attenuate it.
While this setup might be an extreme case, it demonstrates the numerical sensitivity required to adequately handle masing in such models.
Given population inversions have been observed in the atmospheres of solar system planets \citep{96CoMoOr.H2O,09PoGuEl.H2O} effective handling of their numerical complications may be essential for accurately modelling radiative transfer in exoplanets.
Further work is needed to improve the numerical stability of population inversions and maser action in non-LTE radiative transfer models.

\subsection{Cross sections}

\begin{figure*}
    \centering
    \includegraphics[width=1.0\linewidth]{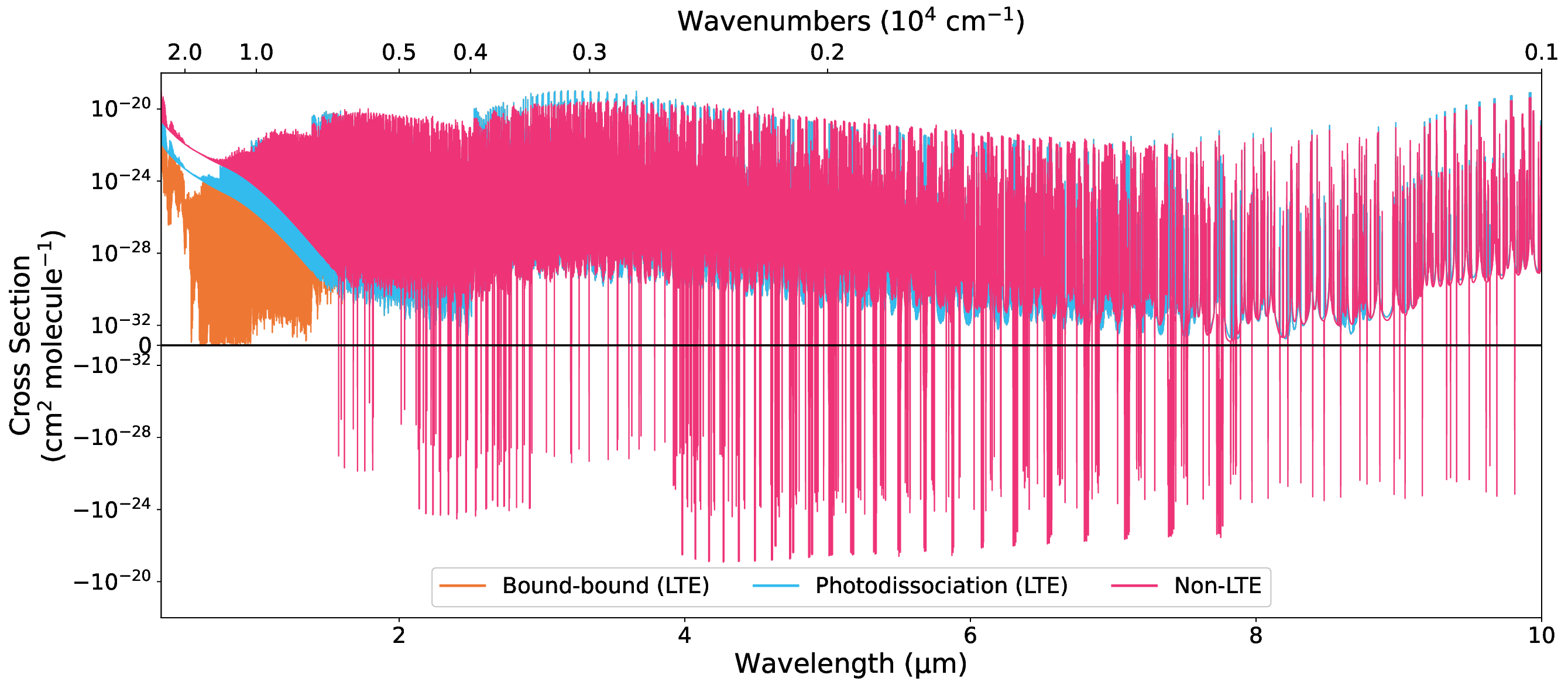}
    \caption{Cross section of OH, computed at the top of a model KELT-20\,b atmosphere at 3375 K and 10$^{-6}$ bar at a resolution of $R = 15\,000$. The lower (orange) cross section is computed from bound-bound transitions in LTE; the middle (blue) cross section shows the LTE case with the inclusion of photodissociation. The upper (pink) cross section is the non-LTE case, where the photodissociation cross section in the near infrared and visible is increased by roughly three orders of magnitude due to the significantly enhanced populations of vibrationally excited states. The upper, positive panel represents absorption while the lower, negative panel represents stimulated emission.}
    \label{fig:oh_xsecs}
\end{figure*}

Cross sections are computed at each layer every iteration using the latest set of populations.
\cref{fig:oh_xsecs} shows a comparison between the absorption cross sections for OH at the top of the model atmosphere in LTE and non-LTE.
There are a considerable number of transitions in the near and mid infrared for which stimulated emission is dominating over absorption due to population inversions.
Though there is not a complete inversion between any of the vibronic levels, the populations are sufficiently similar that inversions arise in transitions involving an upper state with $v^{\prime\prime} = v^{\prime} + 1$ and $J^{\prime\prime} = J^{\prime} - 1$, i.e.: only P-branch transitions.
Inversions do not occur in Q- and R-branch transitions without population inversions between vibrational levels due to the assumption of rotational LTE used here.

\cref{fig:oh_xsecs} shows that the continuum absorption in the visible and near infrared arising from photodissociation is significantly enhanced under non-LTE.
Conversely, the maximal absorption in the mid infrared is reduced due to the inversion of some of the strongest transitions.

\subsection{Observable features}
\label{subsec:observable_features}

\begin{figure*}
    \centering
    \includegraphics[width=1.0\linewidth]{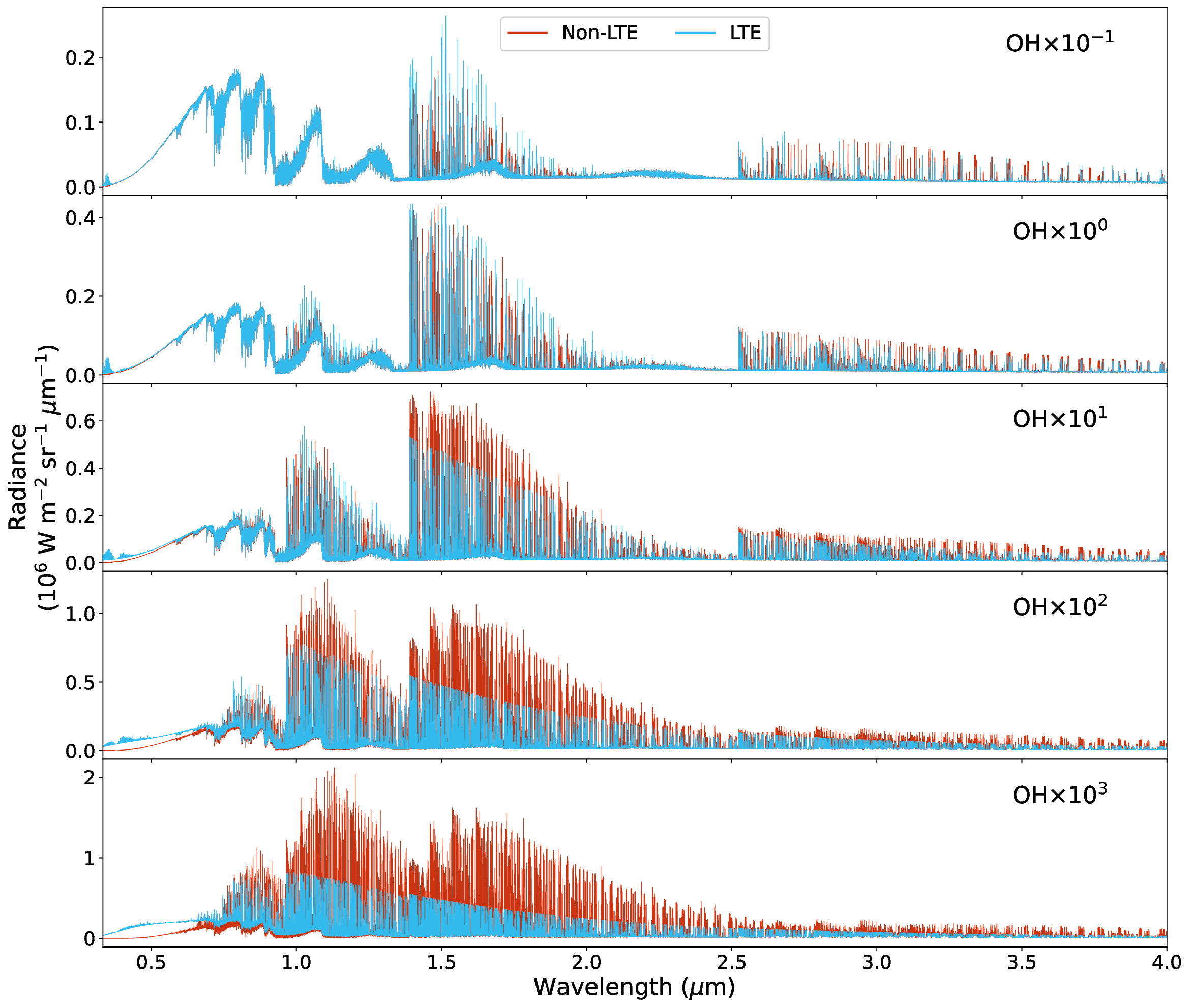}
    \caption{Emission spectra for the model KELT-20\,b atmosphere for the 5 different OH abundance scenarios, each in LTE and Non-LTE.}
    \label{fig:emission}
\end{figure*}

\cref{fig:emission} shows the modelled KELT-20\,b emission spectra computed in both LTE and non-LTE for the five different OH abundance scenarios.
The final emission spectra are computed on a finer grid with a resolution of $R = 15\,000$.
Across all OH abundance scenarios, there are minimal features in the emission spectra longward of 4.0 $\mu$m and the plots shown focus on the strongest bands from the visible to the mid-infrared.
In this region, gaps between emission features become filled in by excited non-LTE hot bands.
In the models with higher OH abundances, the intensity of the fundamental and overtone bands increase significantly, in some cases to more than double that of their LTE equivalent.
These features fall within the wavelength range of the JWST NIRCam instrument.
Alternatively, the features seen in the high-resolutions simulations shown in \cref{fig:emission} should be identifiable with cross-correlation techniques.
In the visible, emission becomes increasingly suppressed with increasing OH abundance, due to the strengthening of continuum absorption bands leading to photodissociation.
These trends are consistent with the changes in the absorption spectra seen in \cref{fig:oh_xsecs} and hence the derived optical depths.
Accordingly, identification of non-LTE effects would be best enabled by observations over broad wavelength ranges.

In the models with the lowest OH abundances, there is a slight decrease in the radiance in the visible, between 0.33-1.0 $\mu$m, in the non-LTE emission spectrum compared to LTE.
As the abundance of OH increases so too does the magnitude of this suppression in the visible.
This is due to the significantly enhanced continuum absorption occurring under non-LTE at these wavelengths, increasing the opacity and hence reducing the observed emission.
Observations of suppressed emission in the visible would need to be accompanied by observations of near-infrared features to constrain the relative magnitude of these effects.

In the OH$\times10^{2}$ and $\times10^{3}$ models, the total radiance in the region between 1.0 - 2.0 $\mu$m in non-LTE is more than double that in LTE at some wavelengths.
In models that only consider molecules under LTE conditions, these enhanced features may be misattributed to a greater abundance of the emitting species.
Alternatively, the shapes of the features in this region change between the two scenarios, with the maximal radiance decreasing approximately linearly across this region in the LTE model.
In non-LTE however, there is a more pronounced dip in the radiance around 1.4 $\mu$m.
This change in the shape of the features may instead be attributed to a different atmospheric species, or in fact result in the failure to identify the source.
This highlights a serious concern for exclusively-LTE retrieval frameworks, which may be failing to fully utilise available molecular data or, more crucially, misattributing observed features.

The discrepancy between the LTE and non-LTE emission is smaller in models run using the collisional rates designed to smooth the transitions between regimes.
This is expected as the larger collisional rates result in a smaller deviation from LTE populations.

\subsection{Photodissociation rates}

\begin{figure}
    \centering
    \includegraphics[width=\linewidth]{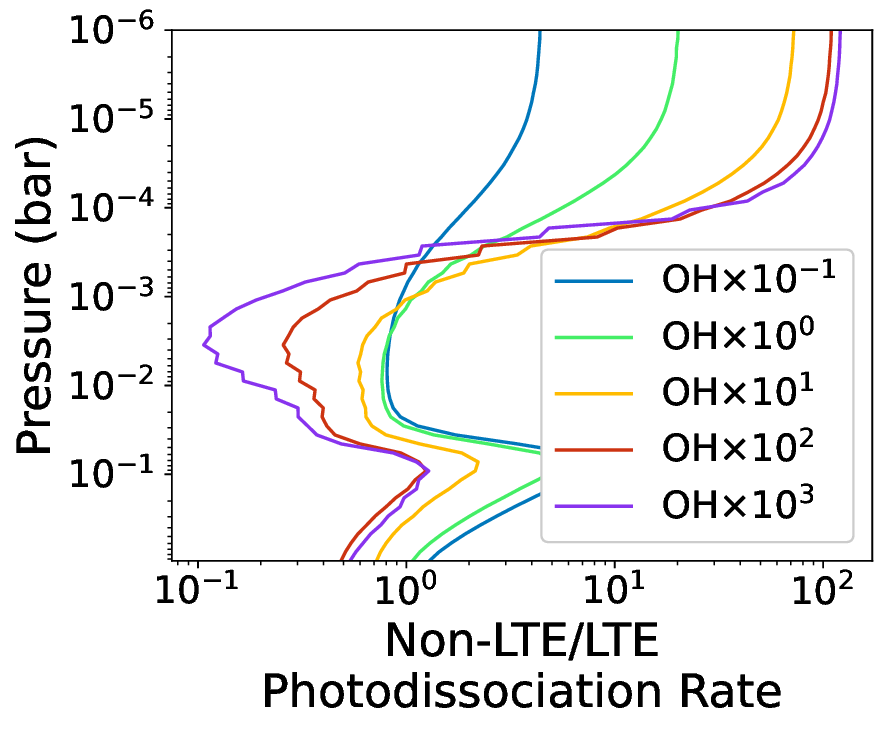}
    \caption{Ratio of non-LTE to LTE integrated photodissociation cross sections as a function of altitude for OH in the model KELT-20\,b atmospheres. Ratios are shown for the 60 layers of the atmosphere ranging from 1 - 10$^{-6}$ bar for which level populations were obtained through solution of the statistical equilibrium equations. The difference between the non-LTE and LTE rates additionally varies as a non-linear function of the OH abundance.}
    \label{fig:oh_photodissociation_rate}
\end{figure}

The total photodissociation cross section was found to change significantly throughout the atmospheres of the non-LTE models, compared to LTE.
Changes in these cross sections directly alter the photodissociation rate of the molecule.
The photodissociation rate is obtained using the photodissociation cross section through the expression
\begin{equation}
    k_{\rm photo.} = \int_0^{\infty} I_{\nu} \sigma_{\nu} d\nu,
\end{equation}
where $I_{\nu}$ is the intensity in the medium and $\sigma_{\nu}$ is the photodissociation cross section.
\cref{fig:oh_photodissociation_rate} shows how the non-LTE photodissociation rate varies throughout the atmosphere compared to LTE.
At the top of the atmosphere, OH photodissociation rates in non-LTE are enhanced by up to a factor of roughly 10$^{2}$ relative to LTE.
These increases are due to the significantly enhanced populations of the vibrationally excited levels of OH under non-LTE, which are more readily photodissociated.

The kink in the ratio of the two photodissociation rates between 10$^{-1}$-10$^{-2}$ bar is likely due to errors in the specific collisional rates, where the $v = 7 - 10$ levels shown in \cref{fig:population_profiles} appear to deviate from LTE while lower vibrational levels remain approximately in LTE.
Increases in the specific collisional rates were found to delay the increase in non-LTE photodissociation rate to lower pressures but in all scenarios, photodissociation cross sections increased in the upper layers of the atmosphere dominated by radiative processes.

\section{Conclusions}

A new program for computing non-LTE opacities and solving radiative transfer within exoplanet atmospheres has been presented.
This code has been used to demonstrate that non-LTE effects should be observable in exoplanet emission spectra and that targeted observations should be able to probe the presence of these effects.
Non-LTE effects have also been shown to impact several areas of atmospheric modelling.
Changes in level populations under non-LTE conditions have been shown to alter a molecule's photodissociation cross section and in turn their photodissociation rates.
Properly accounting for these non-LTE rates will be crucial for accurately modelling the chemical profile of an exoplanet's atmosphere, particularly when accounting for the impact of upper-atmosphere photochemistry.
Packages such as FRECKLL \citep{FRECKLL} are able to dynamically calculate the photodissociation rates throughout an atmosphere when computing chemical equilibria; similar implementations will be necessary to handle chemical equilibria under non-LTE.
As tracers of disequilibrium chemistry are starting to be identified in the atmospheres of exoplanets like WASP-39\,b \citep{23TsLePo.exo}, non-LTE modelling may be key to building an accurate understanding of these environments.
Similarly, increases in the populations of vibrationally excited states will have an impact on retrievals.
Features that would be present in LTE may be observed with a greater strength in emission and lead to the mischaracterisation of molecular abundances.
Significant increases in the strengths of hot bands that would be too weak to be observed in LTE may also be misattributed to features arising from other molecules, if retrieval frameworks only consider LTE opacities.
Retrieval codes will need to account for the fact that under non-LTE conditions the variation in opacity as a function of abundance in non-linear.
These effects will also change the interplay between varying abundance and varying temperature-pressure profiles, which is a known computational challenge \citep{25FoSrKo.nLTE}.
The current implementation of the TIRAMISU code considers a single absorber in non-LTE at a time, though will be extended to allow for the solution of multiple molecules in non-LTE simultaneously.
It has been shown that proper non-LTE treatment of the atmospheric metals can increase the strength of predicted thermal inversions \citep{23FoBiCa.exo} and as such a full treatment of all atmospheric species in non-LTE will be necessary to constrain both the physical and spectroscopic characteristics of an exoplanet atmosphere.
Future work will be carried out to further investigate the ramifications of these effects on exoplanet modelling.

While this work demonstrates the solution for vibronic non-LTE, the rotational populations remain in LTE.
OH has been observed in rotational non-LTE in Earth's atmosphere \citep{18NoPrKa.nLTE} and as such, consideration of rotational non-LTE will be required to comprehensively model exoplanet atmospheres at high resolution.
Nevertheless, the solution of vibronic non-LTE is a considerable step in improving the capabilities spectral modelling and may provide the majority of the contribution to observable non-LTE effects at resolutions such as those available to the James Webb Space Telescope.

While considerable numerical optimisations have been made to make the calculations presented here feasible, the current convergence times for models with a high atmospheric non-LTE fraction are likely too slow for use in retrieval frameworks.
While the majority of the workflow is CPU parallelised, performance increases might be found in a number of calculations through GPU parallelisation.
Application of machine learning techniques to reduce the required number of iterations may also offer considerable performance increases, such as in the implementations of physics-informed neural networks \citep{21MiMo}, convolutional neural networks \citep{22ChPe.nLTE} or extreme learning machines \citep{25TaSiYi.exo} to radiative transfer frameworks.
All such methods obtain accurate and fast predictions of model behaviour and may be useful to expedite non-LTE retrievals.

Inaccuracies in collisional rates affect how rapidly the level populations deviate from LTE in the low-density regions of the atmosphere.
A greater provision of collisional rate data for species of spectroscopic interest, either measured or calculated, is essential for supporting the modelling of non-LTE effects in many astrophysical media.
In hot-Jupiter exoplanets, rate data for collisions between the most common atmospheric constituents, H$_{2}$, atomic Hydrogen and Helium, are essential, particularly at high temperatures.
This also applies to the Hydrogen dominated atmospheres of smaller and cooler exoplanets.
For modelling sub-Neptunes or terrestrial exoplanets, where the bulk atmosphere may be comprised of a wider variety of species, the collisional data required is much broader and more complex.
Unlike in the model atmospheres considered here, consideration of interactions between larger molecules will be more important in cooler atmospheres where these species have a greater mass fraction.
The computation of collisional interactions between such polyatomic species is extremely challenging, particularly for highly excited vibrational states.
Calculations of collisional interactions between two triatomic molecules that only consider rotationally excited states are computationally intensive \citep{25ZoLiZu.HCN}; calcualtions of vibrational quenching rates for such system are not routinely performed but are necessary to support studies of these cooler atmospheres.
While many databases exist that compile collisional rate data, such as BASECOL \citep{jt930}, LAMDA \citep{LAMDA} and EMAA \citep{EMAA}, much of these data are also limited to rotational excitations.
The treatment of collisions between states of vibrational and electronic excitation is needed to support non-LTE modelling in the infrared, visible and even shorter wavelengths.

\section*{Acknowledgements}

This work was supported by the STFC Project No. ST/Y001508/1.

\section*{Data availability}

The ExoMol line lists and opacity data for H$_{2}$O, CO and OH are available via \href{https://www.exomol.com}{exomol.com}.
The TIRAMISU code will be made available soon as part of an open source TauREx plugin.
A development version used to run the models presented here can be found at \href{https://github.com/ExoMol/TIRAMISU}{github.com/ExoMol/TIRAMISU}.

\bibliographystyle{aasjournalv7.bst}
\bibliography{main}

\label{lastpage}
\end{document}